\documentclass[12pt,preprint]{aastex}

\usepackage{epsf}

\def\nbody{$n$-body}
\def\deg{\ifmmode {^\circ}\else {$^\circ$}\fi}
\def\degree{\ifmmode {^\circ}\else {$^\circ$}\fi}
\def\mum{\ifmmode {\rm \,\mu {\rm m}}\else $\rm \,\mu {\rm m}$\fi}
\def\arcsec{\ifmmode ^{\prime \prime}\else $^{\prime \prime}$\fi}

\def\inch{\ifmmode ^{\prime \prime}\else $^{\prime \prime}$\fi}
\def\msunyr{\ifmmode {M_{\odot}~{\rm yr^{-1}}}\else $M_{\odot}~{\rm yr^{-1}}$\fi}
\def\msun{\ifmmode {M_{\odot}}\else $M_{\odot}$\fi}
\def\rsun{\ifmmode {R_{\odot}}\else $R_{\odot}$\fi}
\def\lsun{\ifmmode {L_{\odot}}\else $L_{\odot}$\fi}
\def\mstar{\ifmmode {M_{\star}}\else $M_{\star}$\fi}
\def\rstar{\ifmmode {R_{\star}}\else $R_{\star}$\fi}
\def\tstar{\ifmmode {T_{\star}}\else $T_{\star}$\fi}
\def\lstar{\ifmmode {L_{\star}}\else $L_{\star}$\fi}
\def\md{\ifmmode {M_d}\else $M_d$\fi}
\def\ld{\ifmmode {L_d}\else $L_d$\fi}
\def\ad{\ifmmode A_d\else $A_d$\fi}
\def\ldlstar{\ifmmode L_d / L_\star\else $L_d / L_{\star}$\fi}
\def\rearth{\ifmmode {\rm R_{\oplus}}\else $\rm R_{\oplus}$\fi}
\def\mearth{\ifmmode {\rm M_{\oplus}}\else $\rm M_{\oplus}$\fi}
\def\qdstar{\ifmmode Q_D^\star\else $Q_D^\star$\fi}
\def\kms{\ifmmode {\rm km~s^{-1}}\else $\rm km~s^{-1}$\fi}
\def\ms{\ifmmode {\rm m~s^{-1}}\else $\rm m~s^{-1}$\fi}
\def\mesc{\ifmmode m_{esc}\else $m_{esc}$\fi}
\def\rmin{\ifmmode r_{min}\else $r_{min}$\fi}
\def\rmax{\ifmmode r_{max}\else $r_{max}$\fi}
\def\mmin{\ifmmode m_{min}\else $m_{min}$\fi}
\def\mmax{\ifmmode m_{max}\else $m_{max}$\fi}
\def\rmind{\ifmmode r_{min,d}\else $r_{min,d}$\fi}
\def\rmaxd{\ifmmode r_{max,d}\else $r_{max,d}$\fi}
\def\mmaxd{\ifmmode m_{max,d}\else $m_{max,d}$\fi}
\def\vrad{\ifmmode v_{rad}\else $v_{rad}$\fi}
\def\qz{\ifmmode q_{0}\else $q_{0}$\fi}
\def\qi{\ifmmode q_{i}\else $q_{i}$\fi}
\def\ql{\ifmmode q_{l}\else $q_{l}$\fi}
\def\qs{\ifmmode q_{s}\else $q_{s}$\fi}
\def\rbrk{\ifmmode r_{brk}\else $r_{brk}$\fi}
\def\rdamp{\ifmmode r_{damp}\else $r_{damp}$\fi}
\def\rin{\ifmmode r_{in}\else $r_{in}$\fi}
\def\rout{\ifmmode r_{out}\else $r_{out}$\fi}
\def\tin{\ifmmode t_{in}\else $t_{in}$\fi}
\def\tout{\ifmmode t_{out}\else $t_{out}$\fi}
\def\ain{\ifmmode a_{in}\else $a_{in}$\fi}
\def\aout{\ifmmode a_{out}\else $a_{out}$\fi}
\def\r0{\ifmmode R_{0}\else $R_{0}$\fi}
\def\m0{\ifmmode m_{0}\else $m_{0}$\fi}
\def\M0{\ifmmode M_{0}\else $M_{0}$\fi}
\def\xm{\ifmmode x_{m}\else $x_{m}$\fi}
\def\sigz{\ifmmode \Sigma_0\else $\Sigma_0$\fi}
\def\gyr{\ifmmode {\rm g~yr^{-1}}\else ${\rm g~yr^{-1}}$\fi}
\def\cms{\ifmmode {\rm cm~s^{-1}}\else ${\rm cm~s^{-1}}$\fi}
\def\gcms{\ifmmode {\rm g~cm^{-2}}\else $\rm g~cm^{-2}$\fi}
\def\gcmc{\ifmmode {\rm g~cm^{-3}}\else $\rm g~cm^{-3}$\fi}
\def\atilin{\ifmmode {\tilde{a}_{in}}\else $\tilde{a}_{in}$\fi}
\def\atilout{\ifmmode {\tilde{a}_{out}}\else $\tilde{a}_{out}$\fi}
\def\atil{\ifmmode {\tilde{a}}\else $\tilde{a}$\fi}
\def\ttil{\ifmmode {\tilde{t}}\else $\tilde{t}$\fi}
\def\sqrttt{\ifmmode {\tilde{t}^{1/2}}\else $\tilde{t}^{1/2}$\fi}

\def\iras{{\it IRAS}}
\def\kepler{{\it Kepler}}
\def\herschel{{\it Herschel}}
\def\spitz{{\it Spitzer}}
\def\spitzer{{\it Spitzer}}
\def\wise{{\it WISE}}

\newcommand{\Msunperyr}{M_{\odot}\,{\rm yr}^{-1}}
\newcommand{\Mdotstar}{\dot M_\star}
\newcommand{\gpersqcm}{\rm \,g\,cm^{-2}}

\begin{document}


\title{Rocky Planet Formation: Quick and Neat}

\vskip 27ex

\author{Scott J. Kenyon}
\affil{Smithsonian Astrophysical Observatory,
60 Garden Street, Cambridge, MA 02138 USA}
%\email{e-mail: skenyon@cfa.harvard.edu}

\vskip 4ex

\author{Joan R. Najita}
\affil{National Optical Astronomy Observatory, 
950 Cherry Avenue, Tucson, AZ. 85719, USA}
%\email{email: najita@noao.edu}

\vskip 4ex

\author{Benjamin C. Bromley}
\affil{Department of Physics \& Astronomy, University of Utah,
201 JFB, Salt Lake City, UT 84112 USA}
%\email{e-mail: bromley@physics.utah.edu}

%
%-------------------------- ABSTRACT ----------------------------------
%
\begin{abstract}
We reconsider the commonly held assumption that warm debris disks 
are tracers of terrestrial planet formation. 
The high occurrence rate inferred for Earth-mass planets around mature 
solar-type stars based on exoplanet surveys ($\sim 20$\%) stands 
in stark contrast to the low incidence rate ($\le$ 2\%--3\%) of warm 
dusty debris around solar-type stars during the expected epoch of 
terrestrial planet assembly ($\sim 10$ Myr).  If Earth-mass planets 
at AU distances are a common outcome of the planet formation process, 
this discrepancy suggests that rocky planet formation occurs more
quickly and/or is much neater than traditionally believed, leaving 
behind little in the way of a dust signature.  Alternatively, the 
incidence rate of terrestrial planets has been overestimated or some 
previously unrecognized physical mechanism removes warm dust efficiently 
from the terrestrial planet region. 
%We discuss several observational and theoretical options for eliminating 
%the discrepancy. 
A promising removal mechanism is gas drag in a residual gaseous disk 
with a surface density $\gtrsim 10^{-5}$ of the minimum mass solar nebula.
%Among removal mechanisms, gas drag in a residual gaseous disk with a 
%surface density $\gtrsim 10^{-5}$ of the minimum mass solar nebula is
%the most promising. 

%{\bf JN: I edited the abstract to be more direct and related to the title.
%SK: looks good to me - made a few minor changes }

%We consider the reliability of warm debris disks as tracers of
%terrestrial planet formation.  After examining observational 
%constraints on the frequency of warm dust emission and Earth-mass 
%planets around solar-type stars, we explore theoretical predictions 
%for dust production during the epoch of Earth-mass planet assembly. 
%This analysis confirms that the lack of warm dust emission among 
%5--20~Myr old solar-type stars seriously conflicts with the large 
%frequency of Earth mass planets around their more mature counterparts. 
%After considering observational and theoretical options for eliminating 
%the discrepancy, we conclude that the final stages in the growth of 
%Earth-mass planets occur within a residual gaseous disk having a 
%surface density of $10^{-5}$ of the initial surface density. 
%Gas drag and radiation pressure then remove warm dust particles, 
%reducing emission to well below current detection levels. During
%the next decade, observations with ALMA, HST, JWST, K2, and TESS
%can test our analysis. Theoretical investigations into dust 
%production and the late evolution of protostellar disks enable
%better discriminants among plausible models for the assembly of
%Earth-mass planets.

\end{abstract}



\section{Introduction}
\label{sec: intro}

The conventional picture of
%In the conventional picture, 
terrestrial planet formation begins 
with the growth of 1--10~mm pebbles from 1~\mum\ dust grains 
\citep{chiang2010,birn2010,youdin2010,windmark2012,garaud2013,birn2016}. 
Collisional and collective processes then convert pebbles into 
km-sized or larger planetesimals \citep{youdin2005,johan2007,
youdin2011a,johan2015,simon2016}. 

Various observations support the idea that planetesimals in protoplanetary 
disks grow rapidly (within $\sim$ 1~Myr), including 
(i) radiometric analyses of meteorites \citep{bizzarro2005,
kleine2009,schulz2009,dauphas2011a,dauphas2011b,sugiura2014}, 
(ii) comparisons of the mass distributions of solids in protostellar
disks and known exoplanet populations \citep{najita2014}, 
%%
%% switched order of last two: because the HCN/H2O seems more 
%% directly related to planetesimal formation than HL Tau. 
%%
(iii) the trend in the abundance of HCN relative to H$_2$O as a 
function of disk mass \citep{najita2013}, and 
(iv) coherent structure in the young HL Tau disk, as observed by 
ALMA \citep{alma2015,zhang2015}.  

Once planetesimals form, they quickly ($10^4 - 10^5$~yr) merge 
into much larger protoplanets \citep{weiden1974,weth1980}.  
Over 1--100~Myr, a series of giant impacts among the protoplanets 
then leads to several Earth-mass planets \citep{chambers1998,
chambers2001a,raymond2004,kb2006,lunine2011,raymond2014a}. 
Throughout this period, theory predicts a disk-shaped cloud of 
collisional debris 
that produces an observable infrared (IR) 
excess. These debris disks are expected to serve as signposts of 
ongoing rocky planet formation \citep[e.g.,][]{kb2002b,
zuck2004,kb2004b,raymond2011,raymond2012,lein2015}.

If warm debris is a dependable beacon, it provides a simple way to 
locate sites of ongoing terrestrial planet formation and to measure 
the frequency with which rocky planetary systems form. Compared to 
the challenges of identifying Earth-mass planets from direct imaging, 
microlensing, radial velocity, and transit observations 
\citep[e.g.,][]{gould2006,cum2008,mac2014,burke2015}, measuring the 
magnitude of an IR excess is fairly straightforward and independent 
of viewing geometry \citep[e.g.,][]{carp2009a,carp2009b,kenn2013a,
patel2014}. IR observations are often sufficient to establish the 
temperature and location of the debris \citep[e.g.,][]{lisse2008,currie2011}.
Theoretical models then allow us to use this information to construct 
a window into the planet formation process \citep[e.g.,][]{genda2015a,
kb2016a}.

%% in the interests of making the paper shorter, SK removed two paragraphs
%% we had moved to this section.

%Despite the rarity of warm dust, several systems with warm dust are 
%luminous enough for detailed study
%\citep[e.g.,][]{lisse2008,smith2008,currie2011,olofsson2012,smith2012,
%absil2013,schneider2013,ballering2014,ertel2014,menne2014,defrere2015}. 
%Although current facilities do not enable direct optical/IR imaging of 
%terrestrial disks, interferometric observations suggest several systems 
%have warm dust within 1--2~AU of the central star. In systems with large 
%excesses from 250--400~K dust, extensive analyses of \spitz\ IRS spectra 
%indicate emission from grains with sizes of roughly 1~\mum\ composed of 
%olivine, pyroxenes, silicates, sulfides, and other minerals commonly 
%found in asteroids and meteorites.  According to these investigations,
%the mineralogy of the dust suggests formation in a planet-forming disk.

Here, we examine the reliability of warm debris disks as tracers of
ongoing terrestrial planet formation.  Current observations suggest  
the frequency of warm debris disks around young solar-type stars 
(\S\ref{sec: ddisk}) is much smaller than the frequency of Earth-mass 
planets around older solar-type stars (\S\ref{sec: earths}).  Analytical 
calculations of dust emission during the final phases of planet assembly 
indicate that all popular scenarios of rocky planet formation predict 
detectable amounts of debris for stellar ages of 5--20~Myr (\S\ref{sec: debris}). 
Thus, there is a clear discrepancy between theoretical predictions and 
the observed frequencies of warm debris disks and Earth-mass planets.

In \S\ref{sec: disc}, we consider options for resolving this discrepancy.
After constraining uncertainties in the frequency of Earth-mass planets 
(\S\ref{sec: disc-earth}), dust emission (\S\ref{sec: disc-dust}), and 
theoretical predictions (\S\ref{sec: disc-quick}), we demonstrate that 
if stars retain a residual gaseous disk with a surface density
$\gtrsim 10^{-5}$ of the minimum mass solar nebula (\S\ref{sec: disc-neat}),  
gas drag and radiation pressure will rapidly remove small grains from  
the terrestrial planet region.
This mechanism is plausible given current observational 
limits on gaseous disks among 5--20~Myr old stars (\S\ref{sec: disc-resid}), 
We conclude with a set of suggestions to 
test the possibilities for reconciling theory and observations 
(\S\ref{sec: disc-test}) and a brief summary (\S\ref{sec: summary}).

\section{WARM EXCESSES FROM DEBRIS DISKS ARE RARE}
\label{sec: ddisk}

%{\bf SK: we could eliminate the next paragraph. 
%JN: I trimmed it down a little.}

Although opaque protoplanetary disks surround essentially all newly-formed 
stars \citep[e.g.,][and references therein]{kgw2008,will2011,andrews2015}, 
%With
%bolometric luminosities comparable to the central star, typical protoplanetary 
%disks produce a strong infrared (IR) excess of radiation from near-infrared 
%(NIR) to mm wavelengths \citep[e.g.,][]{adams1987,kh1987,bertout1988}. 
%For optically visible, solar-type pre-main sequence stars, this emission 
%is a factor of 10--100 (40--1000) times the radiation from the stellar 
%photosphere at 10--12~\mum\ \citep[20--25~\mum;][]{kh1995}.  On average, 
the infrared excess emission produced by the disk vanishes roughly 
simultaneously at all wavelengths on a time scale of $\sim 3$\,Myr 
\citep{haisch2001,kenn2009,mama2009,will2011,alex2014}.  The ultraviolet 
excess produced by accretion onto the central star declines on a similar
time scale \citep{hart1998,kenn2009,sicilia2010,ingleby2014}. 
%The dispersion in these time scales is 3--5~Myr. Some 1--2~Myr old stars 
%have little or no disk emission, while some 10~Myr old stars still have 
%substantial disk emission.

Among older, solar-type main sequence stars, roughly 20\% have optically thin 
IR or mm emission from debris disks \citep[e.g.,][]{hillen2008,trill2008,eiroa2013}.  
Typically, this emission is comparable to or less than the stellar flux at 
8--25~\mum\ and a factor of $\gtrsim$ 10 larger than the stellar flux at
longer wavelengths.  To relate excesses to a fractional luminosity, we assume 
the emission arises from a single-temperature, optically thin dust component 
with temperature $T_d$ located at a distance $a$ from a star with luminosity 
\lstar, mass \mstar, radius \rstar, and temperature \tstar:  
\begin{equation}
\left({T_d \over T_\star}\right)^4 \approx \left({R_\star\over 2 a}\right)^{2} ~ .
\end{equation}
The dust to star luminosity ratio is
\begin{equation}
{L_d\over L_\star} = \left({T_d \over T_\star}\right)^4 {A_d\over \pi \rstar^2} ~ ,
\end{equation}
where $A_d$ is the total cross-sectional area of all solid particles.  Assuming 
the grains emit as blackbodies, we can relate $L_d / \lstar$ to the flux ratio at 
any wavelength:
\begin{equation}
{L_d \over L_\star} =
        {\left( e^{h\nu/kT_d} -1\right) \over
                \left(e^{h\nu/kT_\star} -1\right)}
         \left({T_d \over T_\star}\right)^4 
         {F_d \over F_\star} ~ .
\label{eq: ldust}
\end{equation}
For $T_d \approx$ 280~K, \tstar\ $\approx$ 5800~K, and 
$F_d / F_\star \lesssim 1 $ at 12~\mum\ (24~\mum), 
$L_d / \lstar \lesssim 10^{-3}$ ($10^{-4}$).  Debris disks are a 
factor $\gtrsim$ 100--1000 less luminous than the protoplanetary 
disks surrounding T Tauri stars \citep[see also][]{wyatt2008,carp2009a}.

\begin{figure}
\includegraphics[width=6.5in]{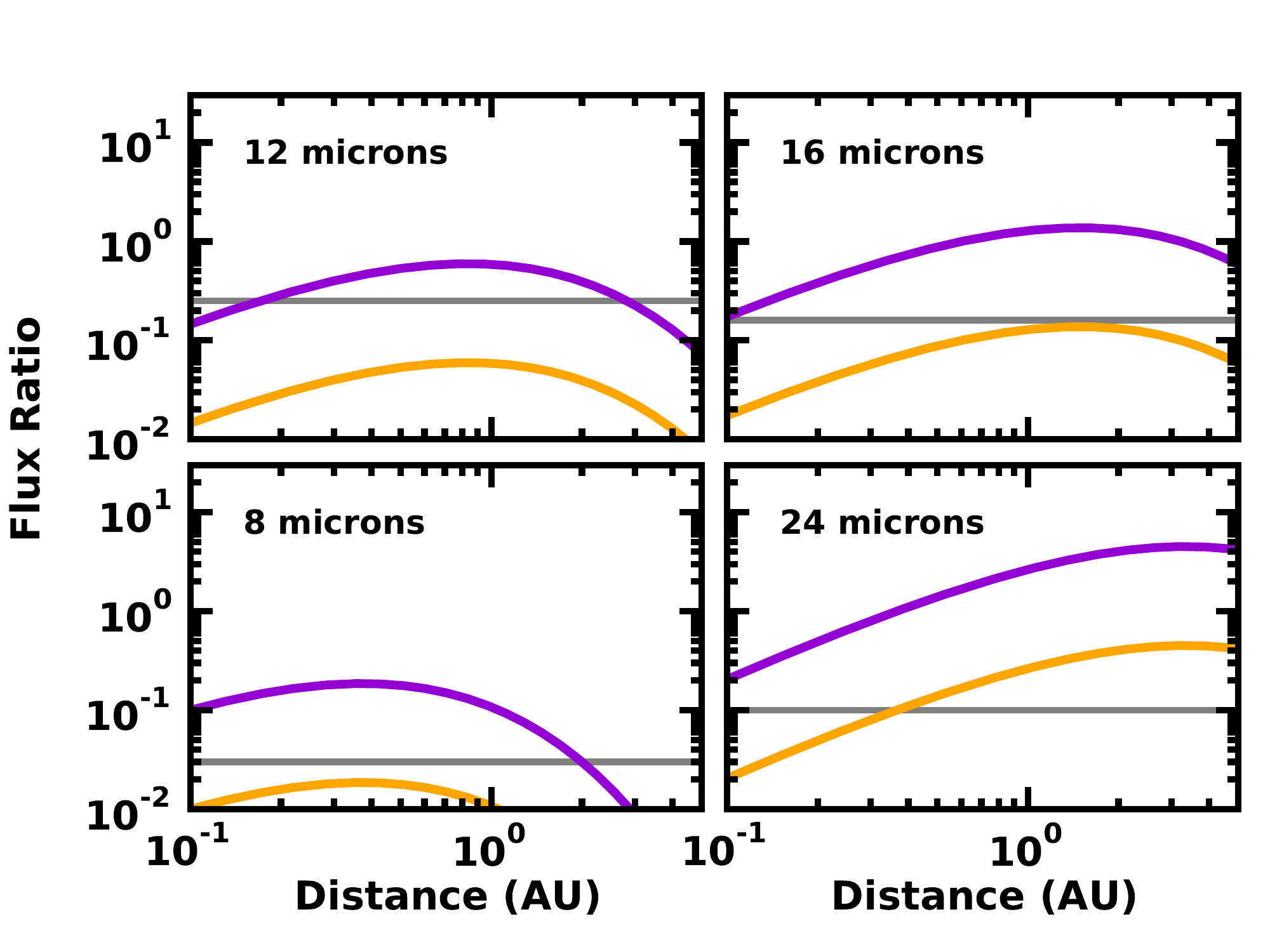}
\vskip 3ex
\caption{
Predicted ratio of dust emission $F_d$ to photospheric 
emission $F_\star$ for material with temperature 
$T_d = 280 (r / {\rm 1~AU} )^{-1/2}$~K and 
$L_d / \lstar = 10^{-3}$ (violet curves) or 
$L_d / \lstar = 10^{-4}$ (orange curves) orbiting a
central star with $T_\star$ = 5800~K. Horizontal grey
lines indicate typical detection limits from \spitz\ 
\citep[8\mum, 16\mum, and 24\mum;][]{carp2009b} 
and \wise\ \citep[12\mum;][]{luhman2012b}.
%
%\citep[e.g.,][]{carp2009a,carp2009b,luhman2012b}.
}
\label{fig: flux-ratio1}
\end{figure}

Fig.~\ref{fig: flux-ratio1} illustrates the variation of $F_d(a)/F_\star$ 
at 8--24~\mum\ for $L_d / \lstar$ = $10^{-3}$ (violet curves) and $10^{-4}$ 
(orange curves). Adopting detection limits from 
\spitz\ \citep[8\mum, 16\mum, 24\mum;][]{carp2009a}
and \wise\ \citep[12\mum;][]{luhman2012b},
$F_d / F_\star \approx$ 0.03--0.25,
it is clear that detecting an excess at 8--12~\mum\ 
from dust at $\sim 1$\,AU requires 
%% $L_d / \lstar \gtrsim 10^{-3}$. 
$L_d / \lstar \gtrsim 3\times 10^{-4}$. 
At longer wavelengths, it is possible to detect warm dust with a 
%%substantially lower luminosity, $L_d / \lstar \gtrsim 10^{-4}$. 
substantially lower luminosity, $L_d / \lstar \gtrsim 3\times 10^{-5}$ 
to $10^{-4}$. 
Because ``cold'' dust ($T_d \lesssim$ 200~K) 
from beyond the terrestrial planet region ($>$1--2\,AU)
can also contribute 
to the observed 24~\mum\ emission, observations at 
shorter wavelengths (e.g., 8--16~\mum) or at longer wavelengths 
(e.g., 70~\mum) are required to constrain $T_d$ and $L_d / \lstar$.

To put an observed $L_d / \lstar$ in perspective, we derive the 
required mass $M_d$ in solid particles. In standard models, the
solids have a power-law size distribution, $N(r) \propto r^{-q}$ 
where $r$ is the radius of a particle and $q \approx$ 3.5. Thus,
$M_d \approx (4/3) \rho A_d (\rmin \rmax)^{1/2}$, where $\rho$ is 
the mass density, \rmin\ is the radius of the smallest particle, and
\rmax\ is the mass of the largest particle \citep{wyatt2008}. For 
material with \rmin\ = 1~\mum, \rmax\ = 300~km, and 
$L_d / \lstar = 10^{-4}$ at $r$ = 1~AU, 
$M_d \approx 3 \times 10^{25}$~g. Thus, an observable excess at 
24~\mum\ requires roughly a third of a lunar mass of solid material 
at 1~AU.

Over the past decade, various surveys suggest a very low frequency of 
warm debris disks among solar-type stars \citep[e.g.,][and references 
therein]{stauffer2005,silver2006,currie2007b,carp2009a,carp2009b,chen2011,
luhman2012b,kenn2013a,clout2014,matthews2014}. 
To illustrate 
%%the efforts to identify robust examples of 
current constraints on the incidence rate of
warm dust, we describe in detail several 
studies using data from \spitz\ and \wise.

The \spitz\ {\it Formation and Evolution of Planetary Systems} (FEPS) program 
surveyed 314 solar-type stars at 3--70~\mum\ \citep{meyer2006,carp2009a}.  
The sample included stars in clusters and the field, with ages ranging from 3~Myr 
to 3~Gyr. Within this group, only 5 have a measurable 16\mum\ excess.
%with $F_d / F_\star \gtrsim$ 0.16 ($L_d / L_\star \gtrsim 10^{-4}$ 
%for dusty material at 1--2~AU; eq.~(\ref{eq: ldust}) and Fig.~\ref{fig: flux-ratio1}).  
%
All five stars also 
have substantial excess emission at 8~\mum, 24~\mum, and 60--100~\mum\ and 
prominent accretion signatures from hot gas close to the central star. Thus, all 
are primordial disks, the gas-rich disks commonly observed in T~Tauri stars 
\citep{silver2006,dahm2009,carp2009a}.

%{\bf SK: JN check below}

%In contrast, the fraction of FEPS sources with 24~\mum\ excesses is fairly large.

All of 
the non-primordial disk excess sources have the modest dust luminosities, 
$L_d/\lstar \lesssim 10^{-3}$, characteristic of debris disks.  
However, few if any sources have obvious emission from warm dust
within 1\,AU of the star \citep{carp2009a}. 
Excluding the primordial disks, 
no source has an excess at $8~\micron$ ($16~\micron$) above 3\% (16\%) 
of the stellar photosphere.
From $R_{24/8}$, the ratio of the 24~\mum\ to the 8~\mum\ flux, 16\% 
of all FEPS sources have robust detections of 24~\mum\ dust emission
$\gtrsim$ 10\% above the stellar photosphere.  Among the 30 stars younger 
than 10\,Myr, 3\%, 7\%, and 20\% of the non-primordial disk sources 
have a detected $24~\micron$ excess at 50\%, 30\%, and 10\% above the 
stellar photosphere.

Detailed fits of model spectral energy distributions for non-blackbody 
grains to the {\it Spitzer} IRS spectra at 8--35\,$\mu$m for excess 
sources of all ages yield average inferred dust temperatures 
$T_d \approx 45$~K to 200~K with a median at 112~K.
Typical inner disk radii range from $a_{\rm in} \approx 2$--3~AU to 40~AU 
with a median at 6~AU.  Adopting a blackbody model for dust emission yields 
similar median values for $T_d$ and $a_{\rm in},$ with somewhat larger ranges 
($T_d \approx$ 50--280~K and $a_{\rm in} \approx$ 1--31~AU). Only a few 
(between 1 and 3) 24~\mum\ excess sources have $T_d \gtrsim$ 200~K and 
$a_{\rm in} \approx$ 1--2~AU, demonstrating that there is little 
obvious evidence for warm dust in the terrestrial zones of FEPS targets.
The excess limits at wavelengths shortward of $24~\micron$ constrain the 
fraction of the $24~\micron$ excess that could be produced by warm dust 
(\S\ref{sec: debris}).

%{\bf JN: suggest to delete the next paragraph?} 
%
%Of all the FEPS targets, only J051111.1+281353 appears to have an excess
%clearly produced by warm dust grains \citep[see Fig.~3 of][]{carp2009a}. 
%The system shows a modest excess relative to a model spectrum at 9--25~\mum, 
%with some hint of a silicate emission feature at 10~\mum. However, the 
%excess cannot be fit by a single temperature blackbody or modified blackbody. 
%Thus, the origin of the excess is uncertain.

%{\bf Joan: The conclusions of Carpenter et al seem somewhat model 
%dependent. If grains radiate as blackbodies, the temp range is 50--280~K
%instead of 45--200~K for the modified blackbody spectrum. They also divide the 
%excess into "warm" and "cool" for luminosity purposes, but this distinction
%is not really warm dust and cool dust - it seems just a convenient notation
%rather than something physical. It can be a little confusing to explain some
%of this analysis so I am not sure how much we should describe. If you want to
%put back some of your description, please do so. 
%}

Although the FEPS study has a high sensitivity to IR excess as a fraction 
of the stellar photosphere, it covers a wide range in stellar age and has a 
relatively small ($\sim 30$) set of sources with ages ($\lesssim$ 10~Myr)
relevant to terrestrial planet assembly.
To complement these results, we examine the {\it Spitzer} and {\it WISE} 
statistics for the well-studied Upper Scorpius association (Upper Sco).
Upper Sco, which is part of the nearby Sco-Cen association
\citep[Sco OB2;][]{preib2008}, 
allows an accurate census of dust emission at ages of 7--12~Myr 
\citep{luhman2012b,pecaut2012,rizzuto2015}, when terrestrial planets 
accumulate most of their final mass 
\citep[e.g.,][]{dauphas2011a,raymond2014a,quintana2016}.

%{\it To focus on the statistics of IR excess sources for a narrower age range 
%within the epoch of terrestrial planet formation}, we consider the
%Upper Scorpius association, which is part of the nearby Sco-Cen association
%\citep[Sco OB2;][]{preib2008}.  This well-studied group of stars 
%allows an accurate census of dust emission at ages of 7--12~Myr 
%\citep{luhman2012b,pecaut2012,rizzuto2015}, when terrestrial planets 
%accumulate 
%most of their final mass \citep[e.g.,][]{dauphas2011a,raymond2014a}.

%Several analyses of \spitz\ and \wise\ data indicate a very small 
%frequency of warm debris disks among solar-type stars in Upper Sco 
%\citep{carp2006,dahm2009,carp2009b,chen2011,luhman2012b}. In a sample of 

Analyses of \spitz\ data alone indicate a very small 
frequency of warm debris disks among solar-type stars in Upper Sco 
\citep{carp2006,dahm2009,carp2009b,chen2011}. In a sample of 
27 K-type pre-main sequence stars with masses of 0.8--1.3~\msun, 
most of the excess sources (7/9) have IR colors consistent with 
primordial disks \citep{carp2006,carp2009b}. Although two 
additional stars have
24~\mum\ excesses consistent with debris disks, both have cool dust 
($T_d \lesssim$ 200~K). Within a much larger sample of 101 M-type stars 
with masses smaller than 0.8~\msun, 9 (17) are debris (primordial) disks. 
The lack of short wavelength excesses among stars with debris disks suggest 
none of these have substantial amounts of warm dust. 

To enlarge the sample of solar-type stars, \citet{luhman2012b} added 
\wise\ data to previous \spitz\ surveys. Among K0--M0 stars with masses
of 0.7--1.3~\msun, 28\% (17/60) have a 22--24\mum\ excess and 13\% (9/68) 
have a 12\mum\ excess at levels $\gtrsim 25$\% of the stellar photosphere. 
There is little evidence for warm debris.
All of the 12~\mum\ excess sources are classified as
``full, transitional, or evolved disks''; none are debris disks.  When
all detection statistics are considered \citep{luhman2012b},
fewer than 3\% of Upper Sco sources have a 12~\mum\ excess from a debris 
disk.  Among the 22--24~\mum\ excess sources without 8~\mum\ or 
12~\mum\ excesses, eight are classified as ``debris or evolved transitional'' 
disks.  For the two stars with large 22--24~\mum\ excesses, the lack of 
an accompanying 8~\mum\ or 16~\mum\ excess indicates that most of the 
dust is cold.  The maximum frequency of warm debris disks in this sample 
is then 10\% (6/60).

Although there are no constraints on dust temperature for the other six stars, 
if the FEPS results are a guide, most contain cold dust. Among FEPS sources 
with 24~\mum\ excesses that overlap in age with Upper Sco (5--20~Myr), roughly 
80\% have cold dust \citep{carp2009a}.  Adopting this scaling for Upper Sco, 
only 1--2 of the six 22--24~\mum\ excess sources contain warm dust.  Thus, a 
more realistic estimate for the fraction of 10~Myr old stars with warm dust 
is $\sim$ 2\%. 

Many other studies also conclude that warm excesses from debris disks 
around solar-type stars are extremely rare \citep[e.g.,][]{moor2009,
stauffer2010,beich2011,smith2011,zuck2011,ribas2012,urban2012,zuck2012,
jackson2012,kw2012,ballering2013,vican2014}.  Although \iras, \spitz, and 
\herschel\ data suggest 10\% to 30\% of solar-type main sequence stars 
have IR excesses \citep[e.g.,][and references therein]{lagrange2000,
eiroa2013,ballering2013}, nearly all sources have color temperatures 
characteristic of cold dust ($T_d \ll$~300~K) at $a \gg 1$\,AU. In 
addition to Upper Sco, \spitz\ data for the young (15 Myr) clusters 
$h$ and $\chi$ Per suggest a small excess fraction of 1\% to 2\% at 
8~\mum\ \citep{currie2007b,clout2014}. Analyses of \wise\ data for 
stars in the solar neighborhood yield similarly small ($\lesssim$ 
1\%--2\%) fractions of sources with a 12~\mum\ excess 
\citep{kenn2013a,patel2014}.  

\section{EARTH MASS PLANETS AT 0.25--1~AU ARE FAIRLY COMMON} 
\label{sec: earths}

In contrast to the small fraction of stars with warm debris disks, rocky 
planets within 1~AU appear to be fairly common companions to solar-type 
stars \citep{youdin2011b,fang2012,batalha2013,foreman2014,silburt2015,winn2015}. 
Although the false positive rate is uncertain, recent attempts to confirm 
{\it Kepler} candidates with ground-based and other space-based observations 
suggest false positive rates ranging from $\sim$ 10\% to 75\% for various 
ranges of planet masses \citep[e.g.,][]{morton2011,santerne2012,fressin2013,
sliski2014,desert2015,colon2015,santerne2016,cough2016,mullally2016,morton2016}.
Here, we focus on a comprehensive analysis of \kepler\ data which
provides a detailed estimate for the occurrence rate of Earth-mass 
planets inside 1~AU. This approach is conservative: the formation of 
lower- and higher-mass rocky planets also produces observable amounts 
of debris. Assuming false positive rates are small, these analyses 
thus yield robust lower limits to the fraction of solar-type stars 
that produce detectable debris at ages of 5--20~Myr.  We return to 
the false positive rate in \S5.

%{\bf sK: do we need Petigura?}
%\citet{petig2013b} report an incidence rate of 0.5 planets per solar-type
%star for planets with radii of 1--4~\rearth\ and orbital periods $<100$~d.  
%For 
%the same range of planet sizes, they estimate that there are approximately 
%half as many planets with orbital periods of 100--400~d as there are within 
%100~d, yielding an incidence rate of $\sim 0.25$. 
%{\it (Delete: Thus, the total incidence
%rate is roughly 0.75 planets per star for periods of $<400$~d.) } 
%In their
%analysis of systems with orbital periods of 200--400~d, the frequency of
%planets with radii of 1--2~\rearth\ is comparable to the frequency of larger
%planets with radii of 2--4~\rearth. Although rates for planets with radii
%smaller than 1~\rearth\ are unknown, the trends at larger radii suggest 
%these smaller planets may be as plentiful as larger planets. 

%\noindent{\bf JN: Possibly delete italicized sentence above, because we're really 
%interested in 100-400d range, not $<400$d. The rate of 0.75 planets 
%per star is for $<400$d.} 

Using Q1-Q16 Kepler data, \citet{burke2015} estimate 0.77 planets 
with radii of 0.7--2.5 \rearth\ and orbital periods of 50--300 days 
($\sim$ 0.25--0.9~AU) per GK dwarf star \citep[see also][and references 
therein]{petig2013b, mullally2015,cough2016}.  
Recent detailed analyses of transiting planets with radial velocity 
measurements suggest most planets with radii smaller than 
1.5--2~\rearth\ have rocky compositions 
\citep[e.g.,][]{weiss2014,marcy2014,buchave2014}. 
Applying these results to their complete \kepler\ samples, \citet{burke2015}
derive a probability of 0.1 for an Earth-mass planet (0.8--1.2~\rearth) 
at 0.86--1.13 AU~and a probability of 0.075 for an Earth-mass planet 
(0.8--1.2~\rearth) at 0.61--0.8~AU.  Taken at face value, these rates 
predict a somewhat larger frequency of planets (per unit area) at 0.7~AU 
than at 1~AU.  Integrating over 0.6--1~AU, the probability is roughly 0.19 
(assuming an intermediate rate for the missing region at 0.8--0.86~AU).
Although \citet{burke2015} do not quote a rate for Earth-mass planets 
within 0.25--0.6~AU, plausible extrapolations of their rates yield a
probability of 0.22--0.25 (0.21--0.23) for an Earth mass planet 
(0.8--1.2~\rearth) within 0.25--1~AU (0.4--1~AU) of a solar-type star.

%\noindent{\bf JN: In the next para, possibly delete reference to 
%Petigura results because 
%they are a lower rate than Burke, which is supposed to be more complete?} 

Unless the false positive rate for the \kepler\ sample of Earth-mass
planet candidates is much larger than 50\%, the roughly 20\% incidence
rate for rocky planets with radii of 0.8--1.2~\rearth\ at 0.25--1~AU 
is much larger than the $\lesssim$ 2--3\% rate of warm debris disks among 
solar-type stars with ages of 5--30~Myr. The discrepancy between the 
apparent formation rate of terrestrial planets and the detection rate 
for terrestrial debris disks is probably much larger than suggested by
these estimates. We anticipate considerable debris from (i) the formation 
of rocky planets smaller than 0.8~\rearth\ and larger than 
1.2~\rearth\ at 0.25--1~AU and (ii) the formation of any rocky planet
at $\lesssim$ 0.25~AU and at 1--2~AU. If these formation channels yield 
a substantial population of rocky planets, the detection rate for 
debris disks is {\it at least a factor of ten} smaller than expected 
from the incidence rates of rocky planets.

\section{DEBRIS GENERATION FROM PLANET FORMATION} 
\label{sec: debris}

%{\bf JN: The paper discusses 4 paths for debris production: 
%(1) assembly of planetesimals; 
%(2) assembly of protoplanets; 
%(3) giant impacts of protoplanets to make planets; 
%and (4) collisions of planetesimals leftover from protoplanet 
%formation. 
%In section 4.1 the first path is tossed out as not being distinguishable 
%from primordial dust, so then we're left with 3 paths. 
%It is useful to spell out that all 3 
%paths are relevant for all 3 scenarios.  
%Probably best to do this in section 4.1?}

To assess the significance of the different detection rates 
for warm debris disks and Earth-mass planets inside 2~AU, we
must estimate the amount and lifetime of debris produced by
terrestrial planet formation. To make this evaluation, we rely
on theoretical estimates from popular scenarios 
that 
%%which 
produce
terrestrial planets on time scales consistent with the solar
system and observations of protoplanetary disks. These scenarios
are variants on models where continued agglomeration of small 
rocky solids yields a stable planetary system 
\citep[e.g.,][]{saf1969,weth1980}.  Solids not incorporated into 
planets provide material for excess emission from dust.

In the {\bf classical picture} developed to explain the Solar System 
\citep[e.g.,][]{saf1969,lewis1972,weiden1974,weth1980}, the process 
starts with a disk of small solids having just enough mass (the 
Minimum Mass Solar Nebula, hereafter MMSN) to reproduce objects 
in the Solar System.  Collisional processes merge small solids into 
km-sized or larger planetesimals, then Mars-mass protoplanets, and 
finally Earth mass planets. During the `giant impact' phase when 
Mars-mass protoplanets merge into Earths, the surface density of 
the gaseous disk is probably $\lesssim$ 1\% of the initial surface 
density; otherwise, gas drag circularizes the orbits of Mars-mass 
objects and prevents giant impacts \citep[e.g.,][]{kominami2002}. 
Throughout the accumulation and ``clean up'' phases, high velocity 
collisions of leftover planetesimals, impacts of intermediate-sized 
protoplanets, and giant impacts of massive protoplanets convert 
15\% to 30\% of the initial mass in solids into debris
\citep{agnor1999,kb2004b,agnor2004,gold2004,raymond2011,genda2015a}.

%{\bf JN: In the classical picture, there's also a period of "clean up" 
%when leftover planetesimals continue to produce debris? 
%Is it possible to introduce that idea and/or the term 
%"leftover planetesimals" in the above paragraph to get that idea out 
%there?} 

%{\bf SK tried to address the above comment in the preceding para.
%JN: looks good.}
%
%{\bf JN: In the next para, is it useful to say that in addition 
%to fragmentation, there is also some (unstated?) inefficiency in 
%converting planetesimals into planets, and the leftover population 
%also produces debris?}

In the {\bf pebble accretion scenario}, dynamical processes within the 
gaseous disk concentrate cm-sized pebbles into large planetesimals with 
radii of 100--1000~km \citep[e.g.,][]{youdin2005,johan2007,youdin2010,
johan2015,simon2016}.  Continued accretion of pebbles and mergers of 
planetesimals eventually produce a set of stable planets 
\citep[e.g.,][]{johan2015,levison2015,chambers2016}. Current investigations 
of pebble accretion ignore the loss of material and debris production 
during mergers of large planetesimals. If fragmentation removes 5\% to 
10\% of the initial mass \citep[e.g.,][]{johan2015}, neglecting this
process has a limited impact on the formation of large planets.  However, 
the mass lost through fragmentation is much larger than the sub-lunar 
mass of solids required to produce a detectable 24~\mum\ excess (\S2). 

%{\bf: SK made changes to next para - ok Joan?}

%{\bf JN: in the next para, Hansen \& Murray 2012 is mentioned 
%along with 5--10 times MMSN. HM12 considered solid masses of 
%30-100 Me within 1\,AU (as also noted on p. 21), 
%a lot more than 5--10 MMSN, because they focused on explaining 
%close-in Neptunes. In their later focus on lower mass planets, 
%HM13 lowered the initial mass to 20 Me within 1 AU, which is at 
%the upper end of 5--10 MMSN.  HM13 might be the more relevant 
%paper to focus on in this paragraph because they 
%were trying to explain rocky planets. Would it make sense to 
%switch the reference to HM12 and HM13 in this paragraph and 
%change 5-10 times MMSN to $\sim 10$ times MMSN?} 

In the {\bf in situ} planet formation scenario, \citet{hansen2012}
suggest that high concentrations of solids inside 1~AU are required 
to explain various properties of the \kepler\ planet 
population\footnote{\citet{volk2015} address aspects of this picture 
in the context of the solar system.} \citep[see also][]{kuchner2004,
chiang2013,hansen2013,hansen2015}.  Starting from an ensemble of 
protoplanets with a solid surface density $\sim$ 10 times larger 
than the MMSN, a series of giant impacts produces a stable system 
of several planets in a few Myr \citep{hansen2013}.  Because the 
calculations start with assumed ensembles of fully formed protoplanets, 
\citet{hansen2013} ignore the inefficiency of assembling solids into 
protoplanets and the debris produced by high velocity collisions of 
leftover planetesimals.  As in recent explorations of pebble accretion, 
the numerical simulations also ignore the loss of material (and 
consequent debris production) in giant impacts. 

For any of these scenarios, tidal torques can drive the radial migration 
of protoplanets through the gaseous disk \citep[e.g.,][]{ward1997,
ida2000b,masset2003,obrien2006b,papa2007,ida2008a,bk2011b,
raymond2014b}. Planets may then form at large $a$ and migrate to 
small $a$. Although typical migration models do not consider 
collisional disruption of small solids inside 2~AU, large-scale
destruction seems likely.  When Earth-mass and super-Earth-mass 
planets migrate inside 2~AU, they excite pre-existing smaller solids 
onto high $e$ orbits \citep[e.g.,][]{armitage2003,bk2011b,walsh2011,
kb2014a}. 
Destructive collisions among these objects then produce copious
amounts of dust and a detectable 24~\mum\ excess 
\citep[e.g.,][]{wyatt2008,jackson2012}.

\subsection{Modes of Dust Production}
\label{sec: debris-epochs}

To develop predictions for the magnitude of the IR excess produced 
during rocky planet formation, we identify likely epochs of dust 
production.  In all of the scenarios discussed above, the growth of 
pebbles and larger planetesimals with $r \approx$ 1~m to 1000~km 
takes place in an opaque gaseous environment where debris mixes 
with pre-existing small particles \citep[see][and references 
therein]{dauphas2011a}.  Because we cannot distinguish `primordial' 
dust from debris, we ignore this phase of dust production. 

As solids grow from planetesimals into protoplanets and then planets, 
there are three modes of dust formation. Mergers of planetesimals into 
protoplanets typically produce modest amounts of debris 
\citep[e.g.,][]{weth1993,kl1999a,weiden1997b}. As protoplanets grow, 
they stir the orbits of leftover planetesimals which never become 
incorporated into a planet.  Among the planetesimals, high velocity 
collisions then begin to produce numerous smaller particles with sizes 
ranging from 1~\mum\ to tens of km \citep{green1978,weth1993,kb2004b,
weiden2010b,raymond2011}.  This debris fuels a collisional cascade, 
where solids are ground down into sub-micron-sized particles which 
are ejected by radiation pressure.  Eventually, giant impacts produce 
larger and larger protoplanets.  Debris from giant impacts adds to 
the debris from collisions of leftover planetesimals and fuels the 
collisional cascade \citep{jackson2012,genda2015a}.

In all three scenarios, the growth of protoplanets, collisions of
leftover planetesimals, and giant impacts generate debris. For
simplicity, we ignore the modest amount of debris generated from
the growth of protoplanets and derive separate estimates for dust 
produced from destructive collisions of leftover planetesimals 
(\S\ref{sec: debris-proto}) and from giant impacts 
(\S\ref{sec: debris-impacts}).  

%{\bf (JN: possibly delete the rest of this paragraph to streamline? }
%{\it In every 
%scenario, the final accumulation of Earth-mass planets involves giant 
%impacts between Mars-mass protoplanets.  Any particle ejected during
%the impact is more likely to collide with another particle of debris
%than with a growing protoplanet \citep[e.g.,][]{kb2004a,raymond2011,
%raymond2012,genda2015a,kb2016a}.  The mass in protoplanets not incorporated 
%into planets thus yields a lower limit on the mass in the collisional 
%cascade. When planetesimals grow into protoplanets, dust production 
%depends on the initial radii of planetesimals \citep[e.g.,][]{kb2010,kb2016a}. 
%For the classical scenario, the mass lost to destructive collisions 
%probably represents an upper limit to the debris in the collisional 
%cascade \citep{kb2016a}.  Thus, our approach allows us to bracket the 
%amount of debris generated from the formation of Earth-mass planets.)} 

\subsection{Debris from Collisions of Leftover Planetesimals}
\label{sec: debris-proto}

%{\bf In the next para, it's mentioned that some calculations yield 
%negligible debris production. That seems to be what we would want. 
%Why aren't those calculations the ones to focus on?}

In classical planet formation theory, semi-analytical and numerical 
calculations 
of collisions among ensembles of rocky planetesimals within 2--3~AU
report debris production ranging from roughly 5\% to almost 50\% of 
the initial mass in solid material \citep{green1978,weth1993,kb2004b,
lein2005,kb2005,chambers2008,weiden2010b,kb2016a}. 
The typical mass in the debris is 10\% to 20\% of the initial mass.  
Significant numbers of destructive collisions begin early, at $\sim$ 
0.01--0.1~Myr, and last until 10--100~Myr \citep[see also][and 
references therein]{morbi2012,raymond2014a,quintana2016}.  

In these calculations, the timing of debris production overlaps epochs
when we expect a significant decay in the surface density $\Sigma_g$ of 
the gaseous disk \citep{hart1998,haisch2001,mama2009,will2011,alex2014}. 
When $\Sigma_g$ is large, gas drag forces debris to spiral into the central 
star. As $\Sigma_g$ declines, the system retains a larger and larger 
fraction of the debris. For young stars with no gaseous disk at ages of 
10--20~Myr, we conservatively estimate that 5\% to 10\% of the initial 
mass in solids is converted into debris.  The formation of an Earth-mass 
planet thus generates 0.05-0.10~\mearth\ in debris, which is 10--20 times 
larger than the mass required to produce a detectable 24~\mum\ excess.

\subsection{Debris from Giant Impacts}
\label{sec: debris-impacts}

To predict debris production from giant impacts, we first consider a
simple analytical model. In this approach, we compare the collision 
energy to the binding energy of a pair of protoplanets \citep[see 
also][]{davis1985,housen1990,davis1990,housen1999}.  
We assume conservatively that all collisions are head-on (impact 
parameter $b$ = 0); more glancing collisions typically yield more 
debris \citep[e.g.,][]{lein2012}. For two protoplanets with mass 
$m$ and radius $r$, the center-of-mass collision energy is 
$Q_c = v_{imp}^2/8$ \citep[e.g.,][]{weth1993,kcb2014,kb2016a}, 
where the impact velocity\footnote{In our approach, both protoplanets 
have the same velocity relative to a circular orbit.  Thus, the impact 
velocity includes a contribution from each one.} is
\begin{equation}
v_{imp}^2 \approx v_{rel}^2 + v_{rel}^2 + v_{esc}^2 ~ .
\label{eq: v-imp}
\end{equation}
Here, $v_{rel}$ is the velocity of each protoplanet relative to a
circular orbit and $v_{esc}$ is the mutual escape velocity of the 
pair at the moment of the collision, 
$v_{esc} = \sqrt{2 G (m + m) / (r + r)} = \sqrt{2 G m / r}$.
The fraction of material ejected during a collision is 
\begin{equation} 
f_{ej} = 0.5 { Q_c \over \qdstar } ~ ,
\label{eq: f-ej}
\end{equation}
where \qdstar\ is the binding energy \citep[e.g.,][]{agnor1999,
benz1999,canup2001,lein2005,lein2009,genda2012,genda2015b}.  
When $Q_c \approx \qdstar$, half the mass of the merged pair 
is ejected to infinity.

Protoplanets typically have relative velocities of 25\% to 75\% of
the escape velocity of the largest protoplanets.  We set 
$v_{rel} \approx e v_K$, where $v_K$ is the velocity of a circular 
orbit and $e$ is the eccentricity.  When protoplanets have masses 
ranging from a lunar mass to an Earth-mass, $e \approx$ 0.1--0.2
\citep[e.g.,][]{chambers2001a,raymond2004,raymond2005,kb2006,kokubo2006,
kokubo2010,chambers2013,quintana2016}.

\begin{figure}
\includegraphics[width=6.5in]{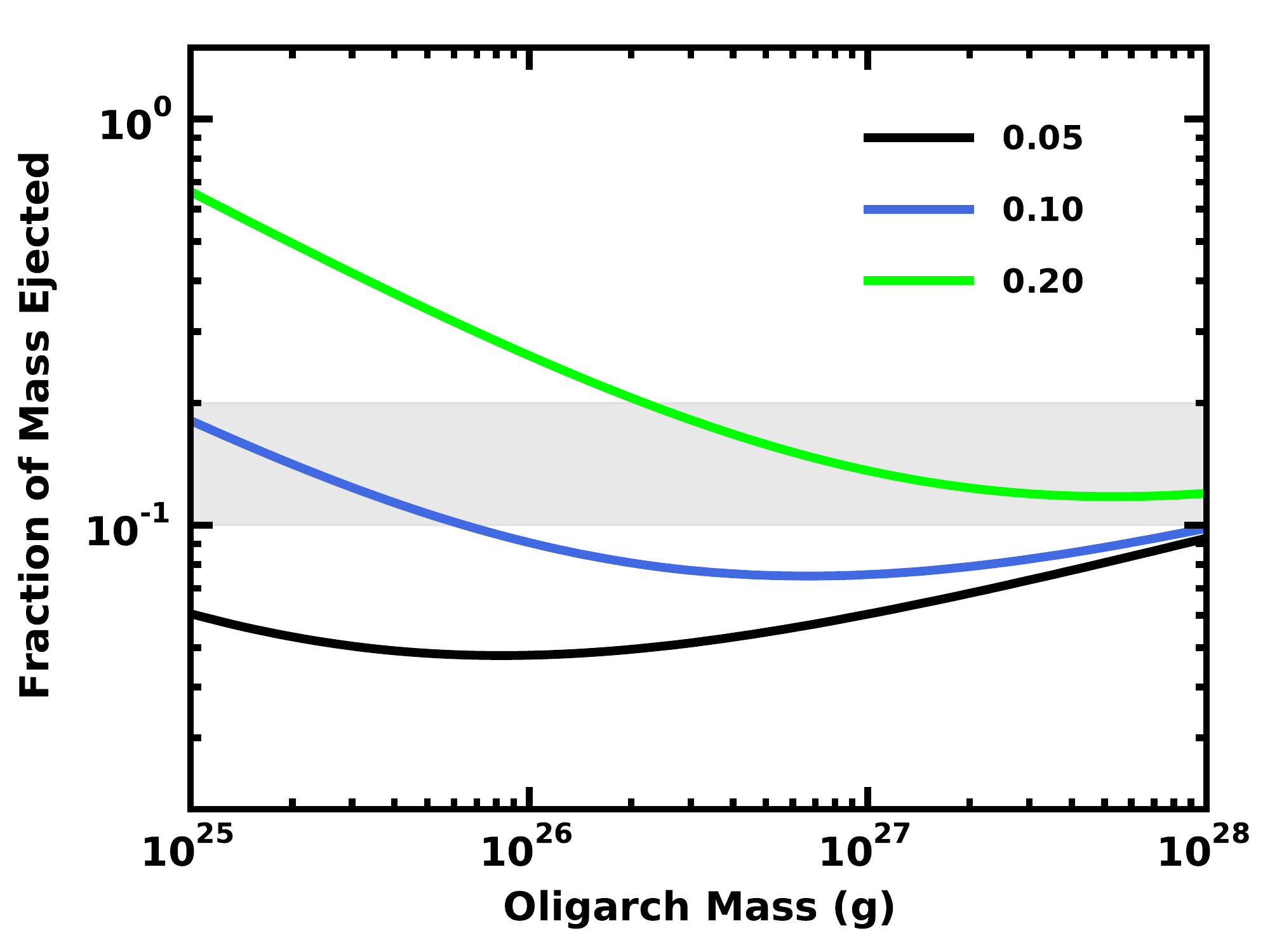}
\vskip 3ex
\caption{
Predicted $f_{ej} = m_{ej} / (m_1 + m_2)$ for collisions
between equal mass oligarchs with $r$ = 1 AU, orbital 
eccentricity $e$ = 0.05 (black curve), 0.10 (blue curve), 
and 0.20 (green curve), and a binding energy 
\qdstar\ appropriate for rocky objects in the terrestrial 
zone. For typical $e \sim$ 0.1, collisions eject roughly 
10\% of the mass of the colliding pair of oligarchs.
The grey bar indicates the range of $f_{ej}$ derived 
from several \nbody\ simulations of protoplanets.
}
\label{fig: f-ej}
\end{figure}

For parameters appropriate for rocky objects at 1~AU (see
\S\ref{sec: app-ann}), collisions between equal-mass protoplanets 
with $e \approx$ 0.05-0.2 eject roughly 10\% of the total mass
(Fig.~\ref{fig: f-ej}). Setting the relative velocity equal to 50\%
of the escape velocity of the protoplanet yields similar results.
Although the accretion history of an Earth-mass object is complicated
\citep[e.g.,][]{kb2006,kokubo2006,chambers2013}, final assembly
requires several collisions of sub-Earth mass objects. If each collision
loses roughly 10\% of the initial mass, the total amount of lost mass
exceeds 0.1~\mearth.  This mass is roughly an order of magnitude larger
than the mass required to produce a detectable debris disk (\S\ref{sec: ddisk}).

Detailed \nbody\ and SPH calculations support this simple estimate 
\citep[e.g.,][]{agnor2004,asphaug2006,raymond2011,genda2012,
stewart2012,jackson2012,chambers2013,genda2015a}.  In SPH simulations, 
dust production depends on $b$ and $v_{rel}$.  Head-on collisions with 
small $b$ and $v_{rel}$, yield little or no dust.  When $v_{rel}$ is 
large or the collision is oblique ($b \gtrsim$ 0.3--0.4), dust production 
is substantial. Averaged over a complete \nbody\ simulation of an ensemble 
of growing protoplanets, protoplanet collisions disperse 10\% to 20\% of 
the initial mass into small fragments. Thus, giant impacts involved in 
the formation of a single Earth-mass planet yield at least 
0.1--0.2~\mearth\ in debris. 

\subsection{Evolution of IR Excess from the Debris of Planet Formation}
\label{sec: debris-an}

To derive predicted detection rates for warm debris disks, we adopt an 
initial disk mass $M_0$ and a model for the time evolution of the total 
mass \md\ and cross-sectional area \ad\ of the debris produced through 
collisions of leftover planetesimals and giant impacts. In a real debris 
disk, stochastic collisions add and remove debris; thus, \md\ and 
\ad\ decrease over the long term and increase and decrease on short 
timescales \citep[e.g.,][]{grogan2001,kb2004b,weiden2010b,jackson2012,
genda2015a,kb2016a}. Instead of following this evolution in detail, we 
consider an analytical model where \ad\ and \md\ decline monotonically with 
time (see \S\ref{sec: app-ann}). With this conservative assumption, we derive 
a lower limit for the expected IR excess from the debris at any time $t$.
If the analytical model predicts much larger IR excesses than observed, 
then more extensive numerical calculations of planet formation will also 
yield much larger excesses than observed.

As outlined in the Appendix (\S\ref{sec: app-ann}), the long-term 
evolution of the debris in the analytical model depends on the surface 
density of solids, the size of the largest object, and the orbital 
eccentricity of debris particles.  The model assumes that collisions
among objects with radii smaller than \rmax\ are completely destructive,
producing debris with particle sizes smaller than \rmax. The cascade
of collisions maintains a power-law size distribution with 
$N(r) \propto r^{-3.5}$ from the smallest size \rmin\ up to \rmax.
Radiation pressure ejects smaller particles. Protoplanets with radii
larger than \rmax\ are ignored.  With destructive collisions for all 
particles smaller than $\rmax$ and ejection of material with 
$r \lesssim \rmin$, the mass in solids steadily declines with time.

To predict the time evolution of dust in the terrestrial zone, 
we consider a disk with initial solid surface density 
$\Sigma_s(a) = \xm \sigz (a/a_0)^{-3/2}$, inner radius \ain\ $\approx$ 
0.1~AU, and outer radius \aout\ $\approx$ 1--2~AU. This inner radius 
lies between the value adopted in some numerical calculations 
\citep[0.05~AU, e.g.,][]{hansen2012,hansen2013} and the 0.25~AU inner 
boundary for the \citet{burke2015} analysis of \kepler\ data.  Our
results are insensitive to the exact value of \ain.
Disks with scale factor \xm\ = 1 and $\sigz$ = 10~\gcms\ at 
$a_0 =$ 1~AU have the surface density of the MMSN.  For 
typical conditions in a disk with several protoplanets, 
\rmax\ $\approx$ 100--1000~km and $e \approx$ 0.1 
\citep{chambers2008,raymond2011,kb2016a}.  If \rmax\ and $e$ 
remain fixed throughout the evolution, the surface density 
declines roughly linearly with time, 
$\Sigma_s(t) \propto (1 + t/t_c)^{-1}$, where 
$t_c \propto \rmax P / \sigz$ is the collision time and $P$ 
is the orbital period.  The relative luminosity of the debris 
disk is then an analytic function of the extent of the disk, 
the initial mass, and time (\S\ref{sec: app-ann}). 

At the start of the evolution, $L_d / \lstar$ depends only on 
\xm\ and the radial extent of the disk (Appendix, eqs.~\ref{eq: linit1}--
\ref{eq: linit2}). The initial dust luminosity scales linearly
with \xm\ and with radial distance as $a^{-3/2}$. When \ain\ = 
0.1~AU and \aout\ = 2~AU, systems with 
$\xm\ \gtrsim 2 \times 10^{-4}$ have dust luminosity larger than
the nominal detection limit, $L_d / \lstar \gtrsim 10^{-4}$.
%The minimum disk mass required to exceed this limit is roughly 
%4 times the mass of the minor planet Ceres\footnote{This minimum
%disk mass is a factor of ten smaller than the mass for a thin ring
%at 1 AU from \S\ref{sec: ddisk}. In a disk extending to 0.1~AU,
%hotter material closer to the star can produce an excess equal 
%to the nominal detection limit with a smaller amount of mass
%than from a thin ring of dust at 1~AU.}

%{\bf JN: in the above, I wonder if it's misleading to talk 
%about how much mass is needed for the initial luminosity to 
%exceed the detection limit, since it's only that bright for a 
%short time? We're more interested in the mass at an age like 10 Myr? 
%How about deleting reference to Ceres and the associated 
%footnote?} 

%{\bf (JN: in sec 2, p. 4, a mass 10 times larger was quoted 
%for the same luminosity. Is this confusing?)} 

%{\bf SK: hopefully the footnote above addresses the confusion.}

Once the evolution begins, the dust luminosity declines on the
local collision time.  For a disk with $\Sigma_s \propto a^{-3/2}$, 
$t_c \propto a^3$. Although most of the mass is in the outer disk 
($M_d \propto a_{out}^{1/2}$), the material closest to the star 
has the largest brightness per unit surface area.  Solids close 
to the star also have the shortest collision time. Thus, the dust 
luminosity begins to decline on the collision time of the inner 
disk, \tin.  The luminosity declines by a factor of roughly two 
on this time scale \citep[see also][]{kw2010}.

On time scales larger than \tin, collisions remove material 
from larger and larger disk radii. At the outer edge of the disk, 
the collision time is \tout.  On time scales between \tin\ and 
\tout, the dust luminosity declines rather slowly with time, 
$L_d / \lstar \propto t^{-n}$ with $n \approx$ 0.3--0.9. Once
the evolution time exceeds \tout, the decline in the dust luminosity 
follows the decline of a narrow ring, $L_d / \lstar \propto t^{-n}$ 
with $n$ = 1.

At late times, the dust luminosity of disks with different 
\xm\ converges \citep[e.g.,][]{wyatt2002,dom2003}. Although the
initial disk luminosity scales with \xm, disks with large 
\xm\ evolve more rapidly than disks with small \xm. This
convergent luminosity depends on \rmax, \xm, and the extent of
the disk.

\begin{figure}
\includegraphics[width=6.5in]{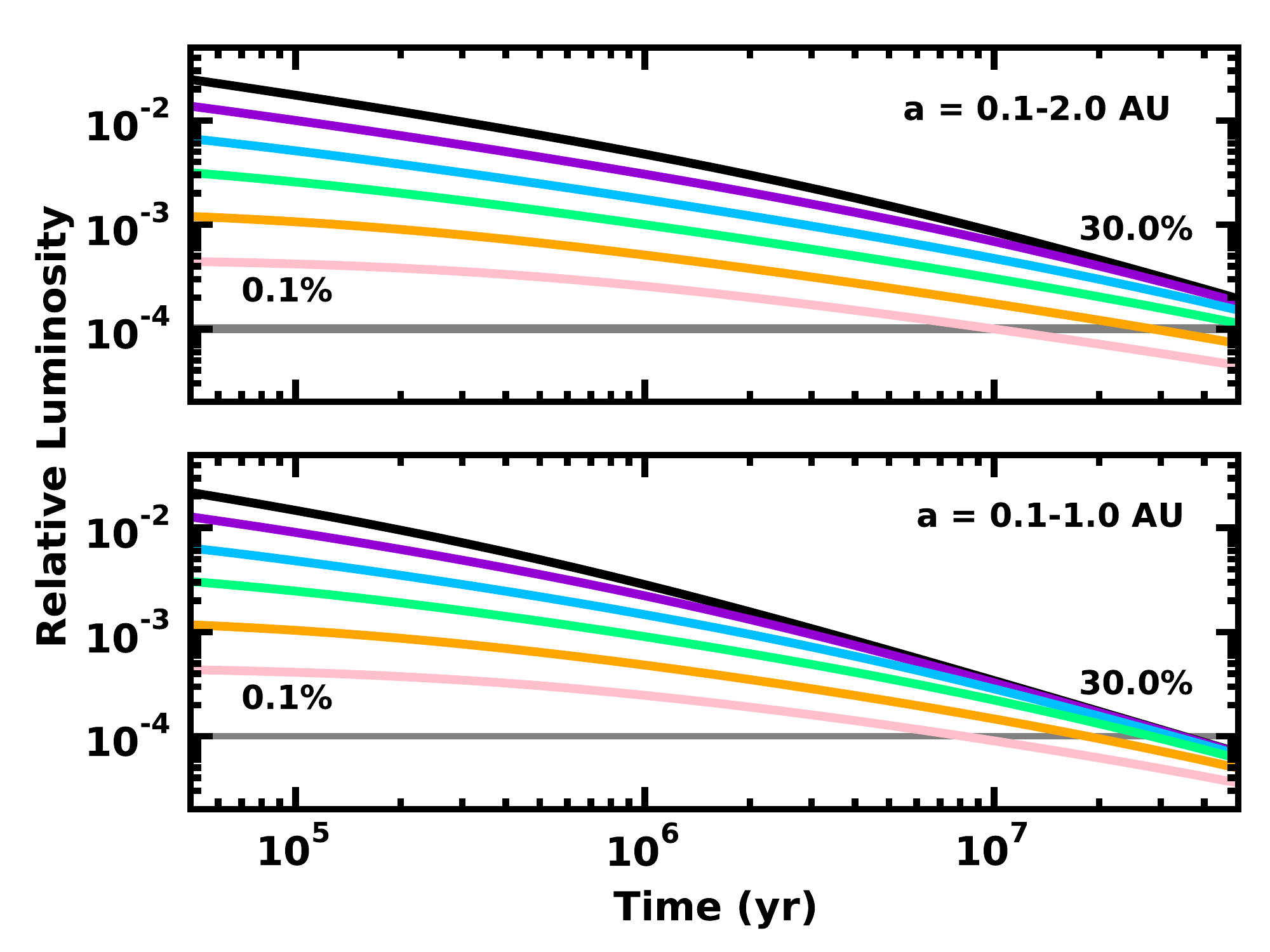}
\vskip 3ex
\caption{
Time evolution of the luminosity ratio $L_d / \lstar$ for 
analytic models of debris disks at 0.1--1~AU (lower panel) 
and at 0.1--2~AU (upper panel). Curves plot results for 
models with \rmax\ = 300~km, $e$ = 0.1, and \xm\ = 0.3 (black),
0.1 (violet), 0.03 (blue), 0.01 (green), 0.003 (orange), and
0.001 (pink).
Predicted luminosity ratios cross the 
nominal detection limit of $L_d / \lstar = 10^{-4}$ at 
10~Myr for models with \xm\ = 0.001 (upper panel) or 
\xm\ = 0.002 (lower panel).
}
\label{fig: lum-ratio1}
\end{figure}

The lower panel of Fig.~\ref{fig: lum-ratio1} illustrates the evolution 
of $L_d / \lstar$ at 0.1--50~Myr for 0.1--1~AU debris disks with 
\rmax\ = 300~km, $e$ = 0.1, and a range of \xm. Disks with \xm\ = 0.3 
(30\% of the MMSN) evolve from $L_d / \lstar \approx 10^{-1}$ at 
$10^3$~yr to $L_d / \lstar \approx 10^{-2}$ at 0.1~Myr to 
$L_d / \lstar \approx 10^{-3}$ at 3--4~Myr. These disks remain much 
brighter than the nominal detection limit until stellar ages of 
30--40~Myr. At early times ($t \lesssim$ 0.1~Myr), the dust 
luminosity roughly scales with \xm; a disk with \xm\ = 0.003 has
a dust luminosity 100 times smaller than a disk with \xm\ = 0.3.  
Because disks with smaller \xm\ have longer collision times, 
they evolve more slowly. It takes a disk with \xm\ = 0.003
roughly 15--20~Myr to decline from $L_d / \lstar \approx 10^{-3}$ 
to $L_d / \lstar \approx 10^{-4}$.

The upper panel of Fig.~\ref{fig: lum-ratio1} shows that larger disks 
are always brighter (eq.~\ref{eq: linit2} of the Appendix).  With a
factor of 8 larger collision time at the outer edge, a disk extending 
to 2~AU declines more slowly than a disk extending to 1~AU. For disks 
with \xm\ = 0.3, the larger disk reaches the detection limit at 100~Myr 
instead of 30--40~Myr. As \xm\ decreases, however, the differences in 
evolution times become smaller. When \xm\ = 0.003, the 0.1--2~AU
disk reaches the detection threshold at 25~Myr instead of 15--20~Myr 
for the 0.1--1~AU disk. 

To derive predicted flux ratios for debris disks, we assume the solids
radiate as blackbodies in equilibrium with radiation from the central
star. For simplicity, \lstar\ = 1~\lsun\, \rstar\ = 1~\rsun, and 
\tstar\ = 5780~K at all times. The disk temperature is then
$T_d$ = 280~$(a /a_0)^{-1/2}$~K. With no analytic solution for
the flux ratio, we divide the disk into a series of annuli, assign a
temperature to each annulus, derive the time evolution of the 
cross-sectional area and emitted flux in each annulus, and add up the
fluxes. In the Appendix, we show that numerical integrations of the
dust luminosity agree well with analytic results.

%{\bf JN: added a reference to 1\% below to connect up with the last 
%sentence of section 4.} 

Fig.~\ref{fig: flux-ratio2} summarizes results for 0.1--2~AU disks with 
\rmin\ = 1~\mum, \rmax\ = 300~km and various \xm. At current sensitivity 
limits, robust 8--12~\mum\ detections of warm dust at 10~Myr require 
massive debris disks with \xm\ $\gtrsim$ 0.3; detections at 8~\mum\ are
strongly favored over those at 12~\mum.
%% \xm\ $\gtrsim$ 0.1--0.3.  
%For these wavelengths, published sensitivity 
%limits strongly favor detection at 8~\mum\ over 12~\mum. 
Warm dust is much easier to detect at 16--24~\mum. 
\spitz\ 16~\mum\ 
%%(24~\mum) 
measurements of $\sim 10$~Myr old 
stars can detect debris disks with \xm\ $\gtrsim$ 0.03. 
%%(0.005).  
The sensitivity at 24~\mum\ is fairly remarkable: 
debris disks with \xm\ as low as 1\% of the MMSN are detectable.
%existing data {\bf are sensitive to} 
%debris disks with \xm\ as small as 0.5\% of the MMSN.

We can use these results to interpret statistics for warm 
debris from \S\ref{sec: ddisk}. Among stars younger than 10~Myr 
in the FEPS sample, 3\%, 7\%, and 20\% of the non-primordial disks 
have a detected $24~\micron$ excess above 50\%, 30\%, and 10\% of 
the stellar photosphere.  From Fig.~\ref{fig: flux-ratio2}, these
results allow us to conclude that $< 3$\% of FEPS sources younger 
than 10 Myr have a warm excess consistent with $\xm \gtrsim 0.1$. 
The 3\% upper limit arises because cold dust beyond 2~AU could 
produce some of the observed $24~\micron$ excess.  Similarly, 
fewer than 7\% of sources younger than 10 Myr have a warm excess 
consistent with $\xm \gtrsim 0.03$.

The FEPS $16~\micron$ excess statistics place a more stringent limit 
on $\xm$.  Because no FEPS source at any age (between 3 Myr and 3 Gyr) 
has a $16~\micron$ excess above 16\% of the stellar photosphere, 
all sources younger than 10 Myr must have $\xm \lesssim 0.05$.  
Similarly, all sources younger than 5 Myr have $\xm \lesssim 0.03$.

For a given \xm, reducing \rmax\ reduces the predicted level of IR 
excess.  Fig.~\ref{fig: flux-ratio3} shows how the 24~\mum\ excess 
varies with \rmax\ for 0.1--2~AU debris disks with \xm\ = 0.1 and 
\rmin\ = 1~\mum. In systems with identical initial masses, swarms 
of particles with smaller \rmax\ have larger initial \ad\ and smaller 
collision times. Although swarms with \rmax\ = 3--10~km initially have 
larger 24~\mum\ excesses than swarms with \rmax\ = 1000~km, they also 
evolve much more rapidly. By 6~Myr (12~Myr), debris disks with 
\rmax\ = 3~km (10~km) reach the nominal detection limit.  Disks with 
\rmax\ = 100--1000~km take 40--100~Myr to reach this limit.

%{\bf The value of \rmax\ must be quite small for the predicted 
%warm excess to reach the 24\mum\ excesses displayed by the 
%brightest 20\% of $\sim 10$\,Myr old stars. 
%If \xm\ is 0.1 and \rmax\ is 10 km, then at 10 Myr 
%$F_d/F_{\rm star} \sim 0.12$ at 24\mum\
%(Fig.~\ref{fig: flux-ratio3}).  
%Approximately 20\% of FEPS sources younger than 10 Myr have 
%an excess brighter than 
%$F_d/F_{\rm star} \sim 0.1$ at 24\mum, a fraction comparable 
%to the fraction of solar-type stars with Earth-like planets. 
%(JN: shall we comment on whether 10 km is a reasonable size?)} 

%Despite the large range in evolutionary time scales for debris disks 
%with \rmax\ = 30--1000~km, swarms with \xm\ = 0.05--0.1 and ages of 
%10--20~Myr are easily detectable at 24~\mum\ ($F_d / F_{\star} \gtrsim$ 
%0.1--0.2). When \xm\ = 0.01, debris disks with $\rmax \gtrsim$ 100~km 
%have IR excesses larger than the nominal detection limit at 24~\mum.

\subsection{Summary} 
\label{sec: debris-summ}

Numerical calculations of planet formation suggest that the agglomeration 
of Earth-mass planets in the terrestrial zone creates copious amounts of 
debris. Current estimates suggest from 15\% to 30\% of the initial mass
is potentially available to produce the debris signature of rocky planet 
formation.  Debris production is probably independent of the mode -- 
classical, {\it in situ}, or pebble -- in which planets form. Unless 
planetesimal formation mechanisms assemble Mars-mass oligarchs directly, 
leaving behind 10\% of the initial mass in debris is inevitable 
\citep[e.g.,][]{kb2016a}. By generating smaller planetesimals, current 
formation paths appear to preclude this possibility \citep{johan2015,
simon2016}. All popular models for the formation of Earth-mass planets 
include a giant impact phase which 
also
generates significant amounts of
debris.

%{\bf In the previous paragraph, xm= 15-30\% as a suggestion. In the next 
%paragraph, 10-20\% is mentioned. Should this also be 15-30\% which lasts 
%for a slightly longer time?}

\begin{figure}
\includegraphics[width=6.5in]{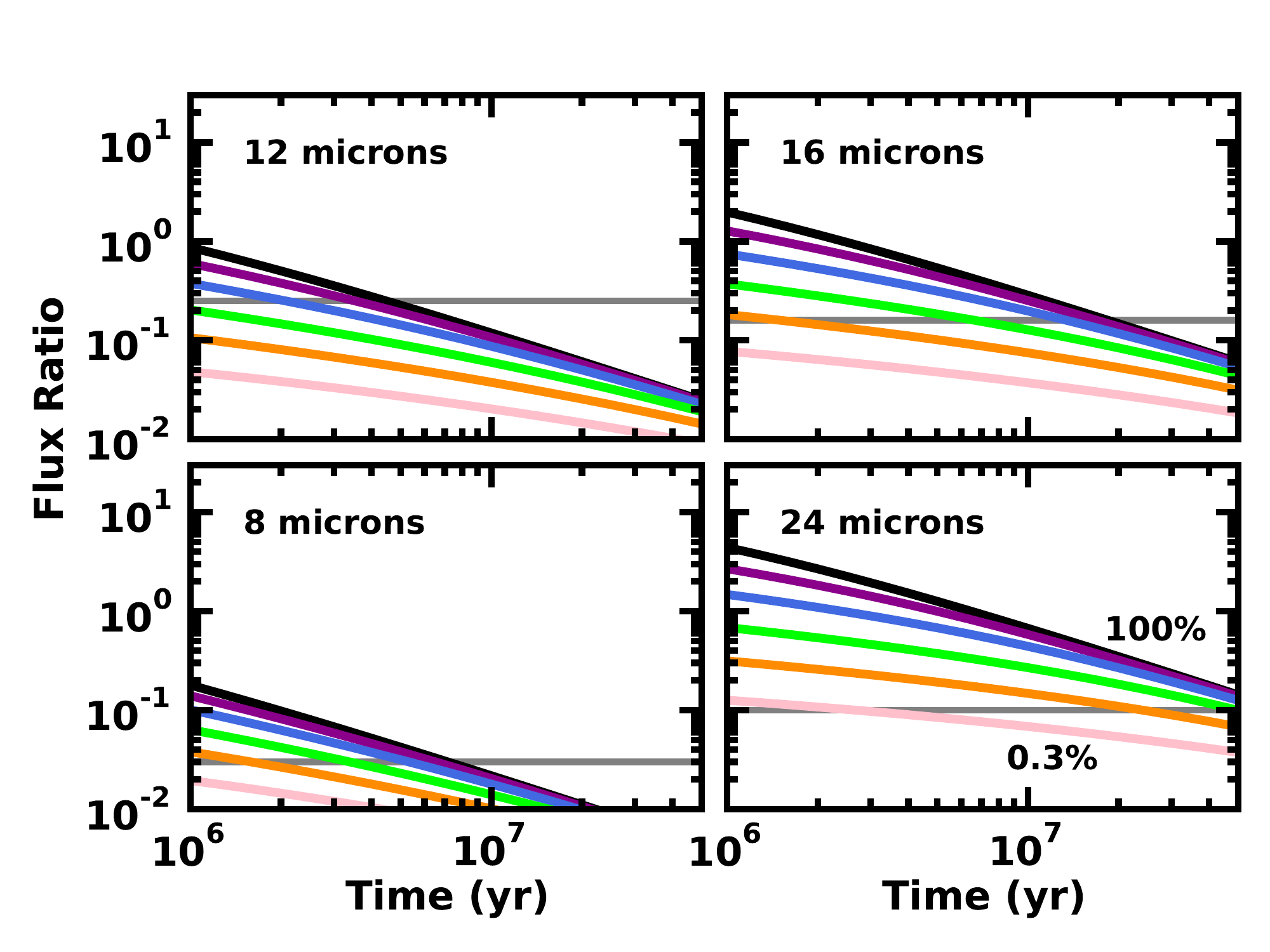}
\vskip 3ex
\caption{
Predicted ratio of dust emission $F_d$ to photospheric 
emission $F_\star$ in an analytic debris disk model with
\rmax\ = 300~km extending from 0.1~AU to 2~AU.  For 
wavelengths of 8--24~\mum\ (as indicated in the legend of
each panel), curves plot results for \xm\ = 1.0 (black 
curves) to \xm\ = 0.003 (pink curves) in steps of 
$\sqrt{10}$. 
Detecting dust from terrestrial planet 
formation at 8--12~\mum\ with existing facilities requires 
conversion of at least 30\% of a MMSN into debris. At
16--24~\mum, as little as 0.3\% of the MMSN is detectable
around stars with ages of 5--20~Myr.
}
\label{fig: flux-ratio2}
\end{figure}

\begin{figure}
\includegraphics[width=6.5in]{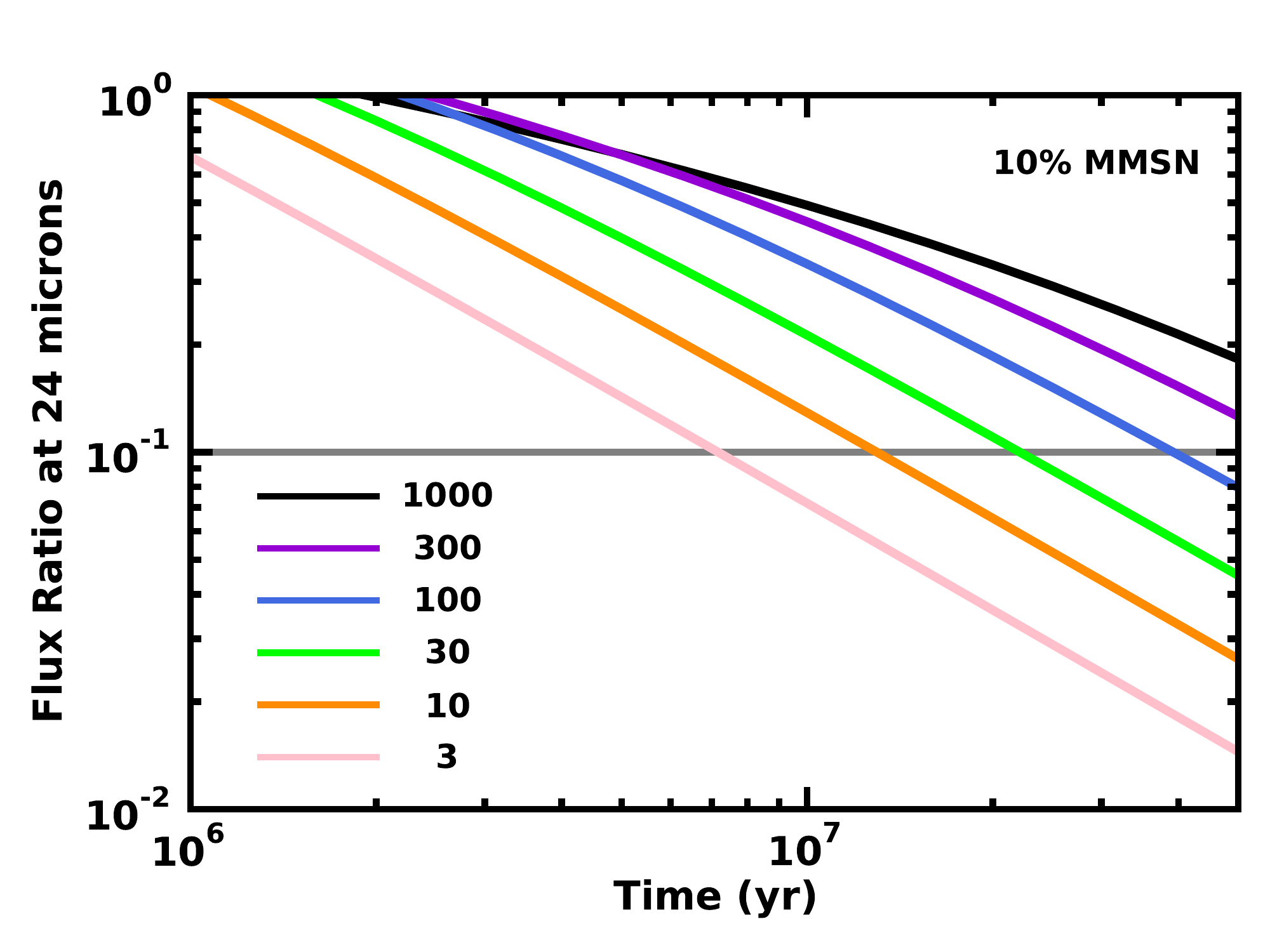}
\vskip 3ex
\caption{
As in Fig.~\ref{fig: flux-ratio2} for 24~\mum\ emission
from disks with \xm\ = 0.1 and \rmax\ = 3--1000~km as 
indicated in the legend. At 10~Myr (20~Myr), disks with
\xm\ = 0.1 and \rmax\ $\approx$ 10--1000~km 
(\rmax\ $\approx$ 30--1000~km) produce detectable amounts
of 24~\mum\ emission.
}
\label{fig: flux-ratio3}
\end{figure}

%Analytical and numerical models suggest 
The debris is long-lived. 
In standard models of planet formation, the debris has a typical
maximum size of 100--1000~km \citep[e.g.,][]{kb2016a}. If the 
debris contains 15\% to 30\% of the initial mass required to assemble 
an Earth-mass planet, then the IR excess from the debris is detectable 
at 24~\mum\ for $\lesssim$ 100~Myr \citep[Fig.~\ref{fig: flux-ratio2}; 
see also][]{raymond2011,genda2015a}. 

%For stars with ages of 10~Myr, 
%detecting debris in the terrestrial
%zone (0.1--2~AU) at 8--12~\mum\ is fairly challenging.  

For stars with ages $\lesssim$ 10~Myr, it is challenging to 
detect debris from terrestrial planet formation with current 
sensitivity limits at 8--12~\mum.  For robust detections, 
recent surveys require an initial solid mass $\gtrsim$ 30\% of 
the MMSN at 8~\mum\ and $\gtrsim$ 100\% of the MMSN at 12~\mum. 
If the formation of several Earth-mass planets converts 15\% to 
30\% of the initial mass into debris, detection at 12~\mum\ is 
very unlikely: the debris simply is not luminous enough after 10~Myr. 
At 8~\mum, the predicted level of debris from the formation of a 
single Earth-mass planet is somewhat smaller than the \spitzer\ limits.
If Earth-mass planets are as common as suggested by 
\kepler\ \citep{burke2015}, the low frequency, $\lesssim$ 3\% 
\citep[e.g.,][]{luhman2012b}, of warm dust revealed by 
8~\mum\ observations of solar-type stars in the Upper Sco 
association is roughly consistent with theoretical predictions.
However, if we include the excesses produced by the formation 
of more massive planets ($\gtrsim 1~\mearth$), the expected 
8~\mum\ excess is larger than \spitzer\ sensitivity limits and 
potentially in conflict with observations.

%{\bf JN: but what about when the debris from the formation of 
%more massive planets is included?  Their formation might 
%produce detectable debris?}

%At 16--24~\mum, the debris from terrestrial planet formation is
%easy to detect. 
If our nominal picture of debris production is correct 
(Fig.~\ref{fig: flux-ratio2}), current 16--24~\mum\ sensitivity 
limits are low enough to detect {\it every} solar-type star 
$\lesssim 10$\,Myr engaged in forming Earth-mass planets.  
For 10~Myr old stars, current 16~\mum\ data should detect 
debris disks with initial masses of 5\% to 10\% of the MMSN;
however, the FEPS survey detected none of these systems.
Similarly, $< 3$\% of FEPS sources younger than 10 Myr have a 
warm excess consistent with initial solid masses $\gtrsim 10$\% 
of the MMSN.
%
%At 24~\mum, current observations should identify 
%debris with only 0.5\% to 1\% of the MMSN. 
%
%Allowing for very 
%generous factor of 2--3 uncertainties in debris production from 
%the assembly of Earth-mass planets, 
%{\bf current 16--24~\mum\ 
%sensitivity limits are low enough to 
%detect {\it every} system $\lesssim 10$\,Myr engaged in} 
%forming Earth-mass planets. 

Based on this analysis, there is a clear discrepancy between the
observed frequency of warm debris from terrestrial planet formation
%%warm debris disks at 24~\mum\ 
($\lesssim$ 3\%, \S\ref{sec: ddisk}) 
and the incidence rate of Earth-mass planets derived from 
\kepler\ ($\gtrsim$ 20\%, \S\ref{sec: earths}). 
Based on Fig.~\ref{fig: flux-ratio2}, the observed frequency 
of 16--24\,\mum\ excess sources and the inferred incidence 
rate of Earth-mass planets imply that terrestrial planet 
formation must leave behind a very small fraction of the initial 
mass in debris-producing solids ($\lesssim 1$\% of MMSN).

\section{DISCUSSION}
\label{sec: disc}

%Our results extend and enhance previous analyses of warm excess 
%rates in solar-type stars. 

The low rate of warm excess among solar-type stars of terrestrial 
planet-forming age was previously noted in the early analysis of 
FEPS \spitz/IRAC results.  
Because terrestrial planet formation should produce a detectable 
warm excess \citep{kb2004b}, \citet{silver2006} interpreted their 
non-detection of warm excesses as possible evidence that 
(i) terrestrial planets form infrequently or 
(ii) warm dust dissipates more quickly than expected
\citep[see also][]{carp2009a}. 
At that time, the rarity of warm excesses was not particularly 
remarkable because the incidence rate of terrestrial planets was
completely unknown. Our more recent understanding that 
Earth-mass planets may be fairly common now highlights the
need to understand why warm excesses are rare.

%With rocky planets now commonly detected around solar-type stars
%(\S\ref{sec: earths}),
%Since that time, it has become clear that rocky planets likely occur 
%commonly in the terrestrial planet region (Sec.\ 3), a 
%development that shifts the focus to other potential 
%explanations for the low rate of warm excesses. } 
%
%%Our conclusion that 
%

In our analysis, the inferred discrepancy between
the fraction $f_d$ of young solar-type stars with 
warm debris 
%
% is much smaller than 
%
and the fraction $f_p$ of mature solar-type 
stars with Earth-mass planets ($f_d \lesssim 0.1 f_p$) relies on our 
understanding of (i) recent \kepler\ planet detection statistics, 
(ii) updated statistics for 8--24~\mum\ emission from disks around 
young stars, and (iii) new developments in planet formation theory. 
In this section, we consider each of these elements in turn to 
identify possible ways to resolve the discrepancy between 
$f_d$ and $f_p$. 
%
%In this section, we begin by considering uncertainties 
%in our estimates for $f_d$ and $f_p$.  
%
We then consider physical 
mechanisms that might reconcile the two frequencies and suggest paths 
to test these ideas.

%The simplest ways to resolve the discrepancy between $f_d$ and $f_p$ 
%involve our interpretation of the \kepler, \spitz, and \wise\ data. 
%In this approach, our understanding of planet and debris formation is 
%basically correct; either (i) we have overestimated the frequency of 
%%%Earth-mass planetary systems by a factor of 2--3 or (ii) we have 
%Earth-mass planetary systems by a factor of $\sim 10$ or (ii) we have 
%overestimated the expected 8--24~\mum\ emission from an ensemble of 
%dust grains.

\subsection{Frequency of Earth-Mass Planets at 0.25--2~AU Around Solar-Type Stars}
\label{sec: disc-earth}

%{\bf SK edited next paras}

When the frequency of false-positives $f_{fp}$ among \kepler\ Earth-mass 
planet candidates is large, we overestimate $f_p$. Current analyses 
infer $f_{fp} \approx$ 10\% to 75\% \citep[e.g.,][]{morton2011,
santerne2012,fressin2013,sliski2014,desert2015,colon2015,santerne2016,
cough2016,morton2016}.  Many studies focus on gas giant planet candidates; 
however, \citet{desert2015} consider whether \spitzer\ observations confirm
transits for Earth-mass candidates with orbital periods $\lesssim$ 100~d.
For individual systems, they derive $f_{fp} \approx$ 1\% to 42\%; the
complete sample suggests a typical $f_{fp} \approx$ 10\%. Curiously,
additional observations confirm planets in several candidate systems 
with large individual $f_{fp}$. Although larger samples are required to 
understand the true f$_{fp}$ for Earth-mass planet candidates, these data 
suggest $f_p$ is not overestimated by a factor of 10 
\citep[see also][]{cough2016,morton2016}.

If multi-planet systems occur commonly and with a large range in orbital 
inclination $\imath$, the true $f_p$ is smaller than our estimates. 
Although transit surveys of such systems may detect only one of two or 
more possible planets, detecting a transiting planet in the system is more 
likely because a planet can be detected from multiple directions.
As a result, the fraction of stars with planets in the habitable zone
$f_p$ can be reduced while maintaining the same average number of 
\kepler\ planets per star overall.

%{\bf SK: perhaps we should have the shout out for Scott in the 
%acknowledgements rather than here? 
%JN: that would be ok. As long as people know that I heard the 
%idea from him. }

%{\bf SK made some changes.
%JN: looks good.}

Although observations do not yet provide a robust estimate of the
frequency of multi-planet systems \citep[e.g.,][and references 
therein]{tremaine2012,fabrycky2014,vaneylen2015,winn2015}, 
comparing results from radial velocity and transit surveys 
provides useful constraints (S.\ Tremaine, private communication). 
%
%
%We can constrain the range of mutual inclinations among planet 
%populations by comparing 
%the multi-planet fractions found from transit and radial velocity 
%surveys (S.\ Tremaine, private communication in his colloquium). 
Transit surveys can detect multi-planet systems only when the 
range of mutual inclinations is $\lesssim$ a few degrees; however,
radial velocity surveys are sensitive to systems with a much 
broader range of mutual inclinations.
The frequency of multi-planet systems from recent \kepler\ catalogs
\citep[0.19--0.20;][]{fabrycky2014,mullally2015,cough2016} is 2/3 of 
the frequency of $\sim$ 0.3 derived from radial velocity measurements
\citep{limbach2015}.  The similarity between the two rates implies 
that \kepler\ does not significantly underestimate the fraction of 
multi-planet systems, i.e., mutual inclinations are small and 
planetary systems are fairly flat. 

While this argument applies to mutual inclinations of planets 
of all masses, additional dynamical considerations can restrict 
the mutual inclinations of systems of Earth-mass planets.
For 
known multi-planet systems, the typical range in $\imath$ is small, 
$\approx$ 1\deg--5\deg\ \citep{tremaine2012,fabrycky2014}. With 
$e \approx$ 2--4 $\imath$ \citep{fabrycky2014}, $e \approx$ 0.035--0.20.
If this range in $e$ is typical of systems with Earth-mass planets, 
theory provides a guide to estimate a possible error in $f_p$ at $a$ 
= 0.25--1~AU \citep[see also][]{tremaine2015}.  From numerical 
simulations, systems of several Earth-mass planets have 
$e \approx$ 0.01--0.05 and $\imath \approx e/2$ 
\citep[e.g.,][]{chambers1998,bk2006,raymond2007a,morishima2008,
chambers2013}.  These systems are dynamically unstable when the 
distance between the apocenter of the inner orbit and the pericenter 
of the outer orbit exceed 10 mutual Hill radii $R_H$, where 
$R_H = (a_{in} + a_{out}) [(m_{in} + m_{out}) / 3~\mstar]^{1/3}$ and
$a_{in}$ ($a_{out}$) is the semimajor axis of the inner (outer) planet
\citep{chambers1996,yoshi1999,fang2013,pu2015,petro2015}. 
When all planets have $e$ = 0.03 ($e$ = 0.10), the maximum number of 
stable Earth-mass planets within 0.4--1~AU is 7--8 (4--5).  The maximum 
reduction in $f_p$ is then a factor of 5--7.

The number of transiting planets in a multi-planet system depends on the
distribution of mutual inclinations \citep[e.g.,][]{tremaine2012,
fabrycky2014}. To make estimates for ensembles of closely packed Earth-mass
planets, we perform a simple Monte Carlo calculation. For adopted dispersions 
in $e$ and $\imath$, $\sigma_e$ and $\sigma_\imath = \sigma_e/2$, our algorithm 
establishes a set of Earth-mass planets with orbital separations of 10 
mutual Hill radii and semimajor axes between 0.25~AU and 1~AU. After randomly 
selecting one of these planets to have impact parameter $b \le$ 1, the code 
chooses random deviates for $\imath$, infers impact parameters for the 
remaining planets, and counts the number of planets with $|b| \le$ 1. For 
Gaussian (Rayleigh) deviates with $\sigma_e \approx$ 0.035 and 
$\sigma_\imath \approx$ 1\deg, closely packed systems have seven (six) 
planets and four (three) transits per system.  When $\sigma_e \approx$ 
0.175 and $\sigma_\imath \approx$ 5\deg, tightly packed systems have 3--4
(2--3) planets and one (one) transit per system.  \kepler\ should detect 
maximally packed Earth-mass planets on roughly circular orbits as 
multi-planet systems, but should fail to identify multiple planets in
highly eccentric systems. Either way, these results suggest $f_p$ might be 
overestimated by factors of 2--4. Factor of ten overestimates are unlikely.

Overall, this analysis suggests that overestimates in $f_p$
are unlikely to reduce the fraction of stars with planets in 
the 0.25-1\,AU region ($\sim 20$\%) to the fraction of stars 
with warm excesses ($<3$\%). Moreover, our estimates are 
conservative. If there is a population of Earth-mass planets 
with $a \approx$ 1--2~AU, then $f_p$ is larger than 20\%. If 
the typical $\sigma_e$ of multi-planet systems is roughly 
4$\sigma_\imath$ instead of our adopted 2$\sigma_\imath$ 
\citep{fabrycky2014}, then the maximum number of Earth-mass 
planets in \kepler\ systems with a single transit is roughly 
two. Thus, uncertainties in $f_p$ cannot reduce the discrepancy 
between $f_p$ and $f_d$.

%{\bf JN: edited paragraph below.} 
%
%{\bf SK put some of this material into paras above - do we need?}
%
%Moreover, reducing the fraction of stars with planets in 
%the 0.4-1\,AU region ($\sim 20$\%) to the fraction of stars 
%with warm excesses ($<3$\%) 
%requires a drastic increase in the number 
%of planets per planet-hosting star. 
%Forming such compact planetary systems requires  
%very large solid masses at small radii. 
%The planetary orbits must be stable over the long term 
%and configured so as to not betray the presence of the additional planets through 
%transits or transit timing variations. 
%It also implies that the outcome of planet formation is 
%essentially bimodal: a system either forms no planets or many planets. 

\subsection{Emission from Small Dust Grains}
\label{sec: disc-dust}

Alternatively, we may have overestimated the expected IR excess 
produced by debris.  To derive this excess (\S\ref{sec: debris}), 
we assume grains with sizes $\gtrsim$ 1~\mum.  In any debris disk 
model, the dust luminosity is sensitive to the size of the smallest 
grains, $L_d \propto \rmin^{-1/2}$. Grains with sizes smaller (larger) 
than our nominal limit of 1~\mum\ are more (less) luminous than 
predicted. Although \rmin\ is clearly a free parameter, observations 
of comets in the solar system and warm debris around solar-type stars 
suggest 1~\mum\ is a reasonable lower limit to the grain size 
\citep[e.g.,][]{lisse2006,lisse2007a,lisse2008,currie2011}.  Thus, 
increasing \rmin\ seems an unlikely way to remove the discrepancy 
between $f_d$ and $f_p$.

We also assume disks with power-law size distributions of small
particles.  In a real collisional cascade, the equilibrium size 
distribution has pronounced waves relative to our adopted
$N(r) \propto r^{-3.5}$ power law \citep{campo1994a,obrien2003,
wyatt2011,kb2016a}.  Adopting an analytic model for the waves 
which matches numerical simulations \citep{kb2016a} yields 
IR excesses a factor of 2--3 larger than the predictions of the 
basic analytical model outlined in the appendix. With this assumption, 
it is possible to detect debris disks with 10\% to 30\% (8--12~\mum),
1\% (16~\mum), or 0.3\% (24~\mum) of the MMSN.

We also assume small grains radiate as perfect blackbodies. In a real 
debris disk, small (1--10~\mum) grains are probably hotter and radiate 
less efficiently than perfect blackbodies \citep[e.g.,][]{bac1993,
dent2000,lisse2006}. If we adopt an extreme model where all grains 
radiate inefficiently (e.g., with emissivity 
$\epsilon \propto (\lambda / \lambda_0)^{-b}$, with 
$\lambda_0$ = 1~\mum\ and b $\approx$ 0.8--1),
the IR excess at 8--24~\mum\ is roughly 50\%
smaller than the blackbody prediction.  At 8--12~\mum, this prediction 
has little impact on our ability to detect excesses around 10~Myr stars: 
blackbody grains are already at or below the nominal detection limits 
for \spitz\ and \wise.  At 16--24~\mum, systems with initial masses of 
10\% (16~\mum) or 1\% (24~\mum) of the MMSN still remain above the 
detection threshold even with an extreme emission model.

Considering the uncertainties in \rmin\ and the radiative 
properties and size distribution of small particles, our 
estimates for the IR excesses of warm debris disks seem 
reasonable.  
Allowing for changes in \lstar\ from pre-main sequence stellar 
evolution \citep[e.g.,][]{bar1998,siess2000,bressan2012,chen2014b,
bar2015} and corresponding variations in dust temperature also 
has little impact on predicted IR excess emission at 8--24~\mum.
Thus, the discrepancy between $f_d$ and $f_p$ remains.

%{\it In a worst-case scenario, it is challenging to detect 
%warm dust with 10\% of the MMSN at 16~\mum, but straightforward to 
%detect dust with 1--3\% of the MMSN at 24~\mum.  Thus, the discrepancy 
%between $f_d$ and $f_p$ remains.} 

%{\bf JN: I had trouble with the last 2 sentences above. 
%As written, the claim is that we are sensitive to small \xm\ with 
%24\mum, so there's a discrepancy. However, small 24\mum\ excesses 
%are actually detected, so on that basis alone, there is no discrepancy.
%I'm not sure which effect led to \xm=0.1 being challenging to detect 
%at 16\mum. Did it get 50\% fainter? If it did, and if the 24\mum\ excess 
%is also 50\% fainter, the stated scenario seems consistent with the 
%observations: $\xm=0.1$, no 16\mum\ excess and 24\mum\ excess down 
%toward 25\% of the photosphere. 
%} 

\subsection{Reconsidering Planet Formation Theory: Quick and Neat}
\label{sec: disc-quick}

If we accept the currently measured frequencies of Earth-mass planets 
and warm debris disks, we must then devise a theory that assembles 
Earth-mass planets with little detectable debris. 

{\bf One way to achieve this goal is if planet formation is quick.} 
If 
Earth-mass planets orbiting solar-mass stars reach their final 
masses during the T Tauri phase, debris ejected from giant impacts 
or collisions of leftover planetesimals
would be 
rapidly mixed with the optically thick 
primordial dust left within the gaseous disk. Instead of producing a 
distinctive `debris disk signature,' the debris simply contributes
to the already large IR excess of the primordial disk. As viscous
evolution and other processes remove the gas, the debris from
Earth-mass planet formation leaves with the gas. 

In the classical theory, planet formation is not quick.
Although protoplanet formation might occur during the T Tauri phase
\citep[e.g.,][]{weth1993,weiden1997a,oht2002,kb2006,kokubo2006,
kokubo2010,kb2016a}, growth then stalls 
\citep[e.g.,][]{weiden1997b,chambers2001a,kb2006,raymond2007a,
chambers2008,chambers2013,raymond2014a,genda2015a,quintana2016}. 
Once the gaseous disk dissipates, giant impacts among Mars-mass 
protoplanets then build Earth-mass planets \citep{kominami2004}, 
generating debris.  The gaseous disk plays no further role in the 
evolution of the debris. 

%{\bf JN: We just need low enough gas to allow collisions between 
%protoplanets ($\lesssim 10^{-2}$ MMSN) and large enough to 
%remove the debris ($\sim 10^{-5}$ MMSN). Does it seem reasonable 
%for disk gas to make a rapid transition from one to the other 
%to avoid a debris signature?} 

Pebble accretion is not intrinsically quick 
\citep[e.g.,][]{chambers2014,johan2015,levison2015}. Recent 
studies suggest the rapid formation of 100--1000~km planetesimals
at $\lesssim$ 1~Myr. These planetesimals rapidly accrete leftover 
pebbles.  Because these models begin with initial masses comparable 
to the MMSN, the giant impact phase begins well after the gaseous 
disk has dissipated. Giant impacts then occur on time scales when 
debris is easily detected.

%{\bf SK made a few changes.
%JN: as in an earlier note on p. 9, 
%the 20-100 Me mentioned in the next paragraph seems to be 
%in conflict with the 5-10 times MMSN that's also mentioned.
%Also changed numbers for fd/f* = 0.1 at 24um.}

The massive disks of solids proposed in the {\it in situ} theory 
($\sim$ 10 times the MMSN)
enable rapid planet formation \citep{hansen2012,hansen2013,hansen2015}.  
Formation times scale inversely with the mass of solids; thus,
protoplanets grow much faster than in the classical or pebble
accretion theories.  Even in a primordial gaseous disk, the 
large number of closely packed protoplanets generates a series 
of giant impacts, leading to multiple Earth-mass planets in 
a few Myr.  The gas can then remove small particles produced 
in giant impacts.  However, the gas probably cannot remove any
leftover planetesimals or other 1--100~km particles generated 
during the giant impact phase. 
%
%%To avoid detection 
To produce a 24~\mum\ excess that is 
%$< 20$\% 
$< 10$\% of the stellar 
photosphere  in a system where the initial mass in solids is 
$\sim$ 10 times the MMSN, the giant impact phase must leave 
behind less than 
%% 0.1\% 
%%0.2-0.4\% 
0.1-0.2\% of the initial mass in 100--1000~km objects
(including any leftover planetesimals). Otherwise, 
the {\it in situ} theory makes too much debris at 10~Myr.

%``Moreover, if protoplanet formation is inefficient in 
%any of the above scenarios, some fraction of the initial 
%mass in solids will be left behind as leftover planetesimals, 
%which can also collide and produce debris. 
%If the leftover planetesimals are 10\%-20\% of the initial 
%solid mass, they will remain detectable for $\sim 10$ Myr?''}

Thus, {\bf another way to satisfy the observational constraints is 
if planet formation is intrinsically neat: Earth-mass planets 
assemble with negligible dust emission.} As described above,
the classical, {\it in situ}, and pebble accretion scenarios are 
not neat enough to match observational constraints.  To isolate 
the appropriate physical conditions for a `neat' scenario, 
we rely on the analytical model outlined in the appendix 
(\ref{sec: app-ann}) and published numerical simulations
\citep[e.g.,][]{kb2016a}. To match current sensitivity limits at 
16~\mum\ and 24~\mum, we must reduce dust production by at least a 
factor of ten (i.e., reducing \xm\ from 15\%--30\% to $<3$\%; 
Fig.~\ref{fig: flux-ratio2}).  In all numerical simulations, dust 
production depends on $v_{imp} / \qdstar$, where $v_{imp}$ is the 
impact velocity and \qdstar\ is the binding energy.  Simple 
dynamics sets $v_{imp}$; changing it significantly is unlikely. 

The binding energy is based on analytical and numerical simulations 
using state-of-the-art equations of state. While increasing 
\qdstar\ seems unlikely, it is worth exploring the required physics.
As a simple comparison, the gravitational binding energy per unit 
mass of a uniform sphere exceeds the \qdstar\ for basalt only when
$r \gtrsim$ 10~\rearth. Thus, the internal degrees of freedom of 
protoplanets are important in setting \qdstar.
While it may be possible to modify these by 
factors of 2--3, order of magnitude changes seem unlikely 
\citep[see also][and references therein]{housen1999,hols2002}.

If dust production cannot be limited, the main alternative is to
reduce the lifetime of the debris. In this approach, a shorter 
debris lifetime lessens the likelihood of detection.  From 
eqs.~\ref{eq: tc0} and \ref{eq: tc0n}, lowering \rmax\ or increasing 
$v_{imp} / \qdstar$ by a factor of ten shortens the lifetime of 
the debris by a similar factor.  Having already ruled out 
factor-of-ten changes in $v_{imp} / \qdstar$, we consider the 
impact of changing \rmax. 

In the analytical model, changing \rmax\ modifies the detectability 
of the debris in two ways. For a fixed mass in debris, the 
cross-sectional area $A_d \propto \rmax^{-1/2}$. Reducing 
\rmax\ therefore increases the initial dust luminosity and 
shortens the collision time. 
As shown in Fig.~\ref{fig: flux-ratio3},  \rmax\ must be quite 
small to reduce the predicted warm excess to the level of the 
24\mum\ excess displayed by the brightest 20\% of $\sim 10$\,Myr 
old stars.  If \xm\ = 0.1 and \rmax\ = 10 km, then at 10 Myr 
$F_d/F_{\rm star} \sim 0.12$ at 24\mum.
Approximately 20\% of FEPS sources younger than 10 Myr have 
an excess brighter than 
$F_d/F_{\rm star} \sim 0.1$ at 24\mum, a fraction comparable 
to the fraction of solar-type stars with Earth-like planets. 

%For the calculations summarized in 
%\S\ref{sec: debris-an}, matching the 
%24~\mum\ sensitivity limit 
%with 10\% of the MMSN in dust requires 
%%
%%$\rmax\ \lesssim$ 3~km 
%%
%{\bf $\rmax\ \lesssim$ 30~km (JN: I think we just have to get 
%down to fd/fstar = 0.1-0.2, because 20\% of 
%stars have a 24\mum\ excess above 0.1.)} 
%instead of 30--1000~km. The initial disk luminosity with the smaller 
%\rmax\ is roughly 
%{\bf XX~times larger.} 
%%
%%10~times larger.  

Although this solution seems attractive, it has several major
drawbacks. Numerical simulations of giant impacts already yield
24~\mum\ fluxes much larger than observed \citep{genda2015a}; 
generating a smaller likelihood of detection at 10 Myr age
in exchange for a much larger initial 
luminosity is probably a poor trade. For dust generated from 
planetesimals, numerical simulations suggest that collisional 
damping of small particles maintains large dust luminosities 
far longer than suggested by the analytical model \citep{kb2016a}. 
Increasing the population of small particles by lowering 
\rmax\ probably exacerbates this problem. 

\subsection{Neat Planet Formation in a Remnant Gas Disk}
\label{sec: disc-neat}

If we cannot substantially alter the observed frequency of 
terrestrial planets, models for the assembly of terrestrial 
planets, or the amount of debris generated by terrestrial planets, 
some process must remove small particles with $r \lesssim$ 
0.1--1~mm from the terrestrial zone.  Possibilities include 
interactions with the stellar radiation field, the stellar wind, 
or a remnant gaseous disk.  If one of these processes removes 
small particles faster than the collisional cascade produces them, 
the predicted IR emission from warm dust is reduced by the requisite 
factor of $\sim$ 10, 
%%The predicted 24~\mum\ excess then falls at 
%%or below current detection limits, 
eliminating the discrepancy between $f_d$ and $f_p$.

Compared to aerodynamic drag from a remnant gaseous disk, other 
mechanisms probably have a limited role.  Observations suggest 
radiation pressure may remove particles smaller than 1~\mum, but 
larger particles are robustly detected \citep[e.g.,][and references 
therein]{lisse2008,currie2011,matthews2014}. 
For $L_d / \lstar \gtrsim 10^{-7}$, Poynting-Robertson drag is 
ineffective \citep{wyatt2008}. In the inner Solar System, stellar 
wind drag is roughly 30\% as effective as Poynting-Robertson drag 
in removing small particles \citep{gust1994}. Observations of the
youngest solar-type stars suggest mass loss rates $\lesssim$ 10
times the mass loss rate $\dot{M}_\odot$ of the current Sun 
\citep[][and references therein]{wood2014}. Theoretical studies 
predict mass loss rates up to 100 $\dot{M}_\odot$ \citep[e.g.,][]
{cohen2014,airapetian2016}. Both of these estimates fall well
below the 1000 $\dot{M}_\odot$ required for stellar wind drag to
remove small particles rapidly when $L_d / \lstar \gtrsim 10^{-4}$.
Finally, the sputtering rate for small particles is too small 
compared to the collision rate \citep{wurz2012}.

In some circumstances, photophoresis\footnote{The photophoresis 
mechanism exploits the temperature gradient across an illuminated 
grain embedded in a low pressure gas. Momentum exchange with the gas
results in the grain moving away from the radiation source.} drives 
small particles radially outward through an optically thin gaseous 
disk \citep[e.g.,][]{krauss2005,hermann2007,cuello2016}.  When 
photophoresis is effective, the time scale for radial drift is 
comparable to (and sometimes shorter than) the time scale for 
radial drift due to gas drag. However, the drift time is 
sensitive to the ratio of the heat conductivity to the particle 
asymmetry factor, which is uncertain \citep[e.g.,][]{vonborst2012}.  

Here, we focus on the impact of gas drag, where the physical 
properties of small particles are less important. Our goal is
to identify a range of $\Sigma_g$ for the gaseous disk that 
allows giant impacts (0.1--1\% or less of the MMSN) and removes 
0.1--1~mm and smaller particles on time scales shorter than the 
collision time. 

\begin{figure}
\includegraphics[width=6.5in]{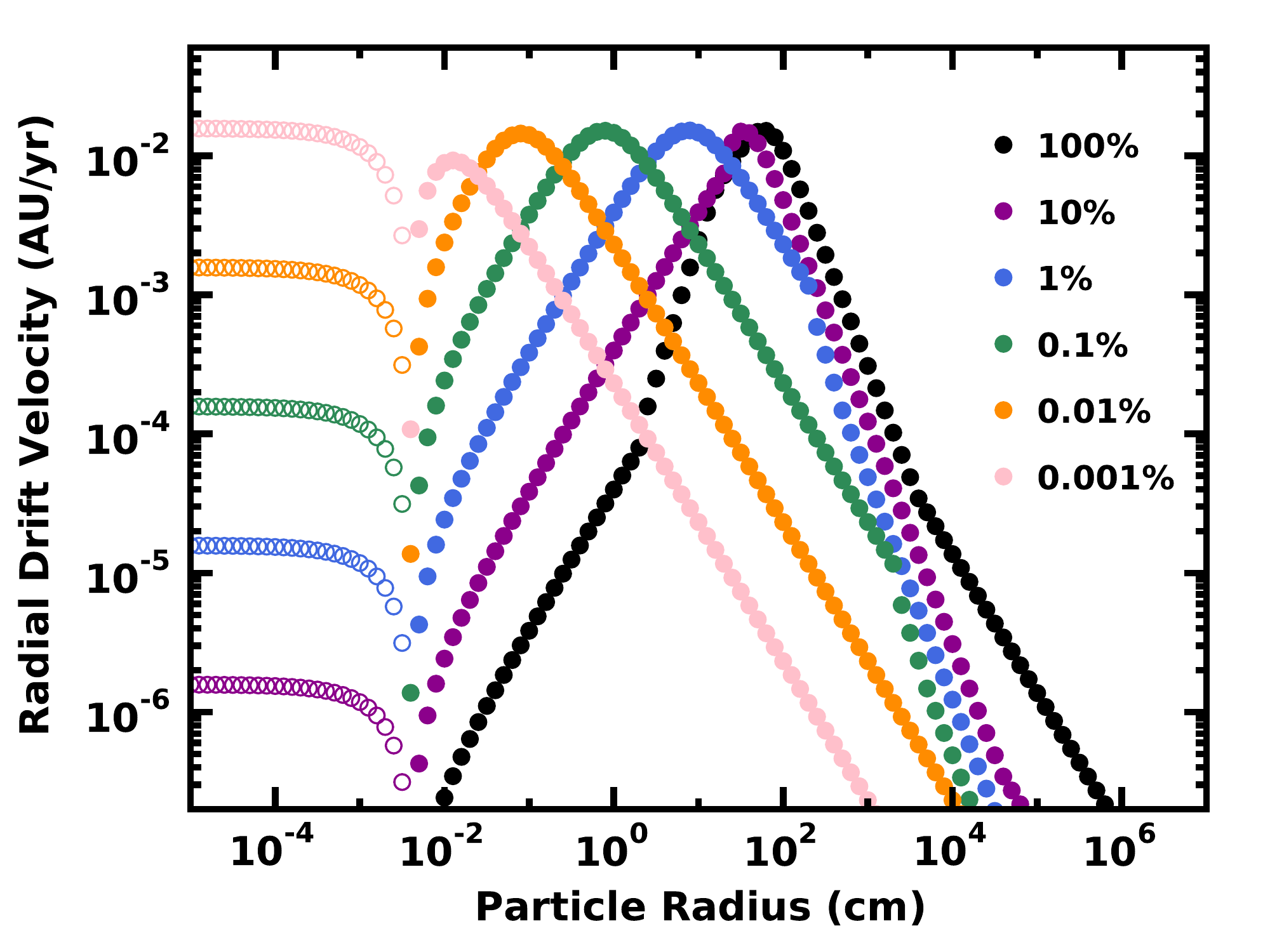}
\vskip 3ex
\caption{
Radial drift velocity of particles at 1 AU in a gaseous disk
with $\Sigma_0$ = 2000~\gcms, $T_0$ = 278~K, $p$ = 1, $q$ = 0.5, 
$H_0$ = 0.03, $s$ = 1/8, $\gamma$ = 1.4, and $\mu$ = 2.0.  Filled 
(open) circles indicate inward (outward) drift.  The legend indicates 
the local surface density relative to the MMSN. At the maximum drift
velocity for particles with $\tau_s \approx$ 1, 50--100~yr lifetimes
are shorter than collisional lifetimes of 1000~yr for debris disks 
with $L_d / \lstar \approx 10^{-4}$. When the surface density is
less than roughly 0.003\% of the MMSN, lifetimes for very small 
particles are also shorter than the collisional lifetime.
}
\label{fig: drag1}
\end{figure}

Within any gaseous disk, particles weakly bound to the gas rapidly 
drift inward when their `stopping time' ($t_s = m v_d / F_d$, 
where $v_d$ is the drift velocity and $F_d$ is the drag force) 
is comparable to their orbital period $P$ \citep{ada1976,weiden1977a}.  
At 1~AU, the shortest drift time is 50--100~yr (Fig.~\ref{fig: drag1}), 
shorter than the
typical collision time of 1000~yr for 1~\mum\ to 1~cm particles.  
In an optically thin disk, radiation pressure drives small,
weakly coupled particles to large $a$ where the particles are
colder and have much smaller dust luminosities \citep{take2001}. 
Effective radial drift thus eliminates particles that produce 
an IR excess\footnote{In a dense gaseous disk, interactions with
the gas rapidly circularize the orbits of small particles 
\citep[e.g.,][]{ada76,weiden1977a}. During a collisional cascade 
in a low density disk, gravitational stirring by protoplanets
raises $e$ faster than gas drag lowers $e$ \citep[e.g,][]{kb2004b,
raymond2011,genda2015a}. Thus, we ignore this process.}.

To quantify radial drift at 1~AU in a low density gaseous disk, we 
combine the approaches of \citet{weiden1977a} and \cite{take2001}.
As outlined in the Appendix (\S\ref{sec: app-gd}), we solve for 
the radial and azimuthal velocity of particles relative to the gas 
in a protostellar disk with standard relations for the surface 
density, midplane temperature, vertical scale height, gas pressure,
thermal velocity, and other physical variables \citep[e.g.,][]{kh1987,
chiang1997,raf2004,chiang2010,youdin2013,armitage2013}.

Fig.~\ref{fig: drag1} illustrates the impact of a residual gas disk 
on the radial drift velocity of particles as a function of size at 
1 AU under MMSN-like conditions.  For this example, a disk with the
surface density of the MMSN has $\Sigma_g$ = 2000~\gcms, $T$ = 278~K, 
and vertical scale height $H$ = 0.03~AU. 
Filled circles indicate inward drift; open symbols indicate outward 
drift.  Other choices for these parameters -- e.g., $T$ = 150--300~K 
and $H$ = 0.02--0.10 -- change the maximum drift velocity and the 
particle size for this maximum drift by 25\% to 50\% but do not change 
the overall trends. 

When the disk has $\Sigma_g$ comparable to the MMSN (Fig.~\ref{fig: drag1}, 
black symbols), particles with $r \approx$ 50~cm have $\tau_s \approx$ 1 
and the largest drift velocity. Inflection points in $\vrad(r)$ occur 
when particles enter different drag regimes (e.g., Epstein, Stokes, 
quadratic).  For particle sizes $r \approx$ 30--40~\mum, 
outward radiation pressure balances inward gas drag;
these particles do not drift. Smaller particles drift 
outward\footnote{MMSN disks are probably optically thick.  Small 
particles in the optically thin upper layers drift outward, while 
those in the optically thick midplane are unaffected.} with maximum 
velocities slightly smaller than 0.1~\cms

As the surface density of the gaseous disk declines, smaller particles 
drift more rapidly through the gas.  Although the peak in \vrad\ shifts 
to smaller sizes, the maximum drift velocity is always roughly 8000~\cms. 
The balance between radiation pressure and gas drag remains fixed 
for 30--40~\mum\ particles. However, in disks with lower $\Sigma_g$, 
smaller particles have larger stopping times. With a weaker drag 
force, radiation pressure drives these particles outward at larger 
velocities. Once $\Sigma_g$ is roughly 0.001\% of the MMSN, the outward 
drift velocities of $r \lesssim$ 30~\mum\ particles surpass the 
maximum inward drift velocities of 100~\mum\ particles.

To judge the impact of radial drift on the collisional cascade, we compare
the drift time to the collisional time. For disks with surface densities 
of 0.001\% of the MMSN, 1--10~\mum\ and 100~\mum\ particles at 1~AU drift 
inward or outward on 50--75~yr time scales. In less (more) tenuous gas 
disks, particle lifetimes are longer (shorter; Fig.~\ref{fig: drag1}). 
For comparison, in a collisional cascade at 0.1--2~AU with 
$L_d / \lstar \approx 10^{-4}$, the collision time is roughly 1000~yr.  
In disks with $\Sigma_g \approx$ 0.001\% of the MMSN, radial drift from
radiation pressure and gas drag 
removes particles smaller than 100 \mum\ in $< 100$ yr, 
rapidly depleting the disk of the 1--100~\mum\ particles that
comprise 90\% of the surface area.  Gas drag also rapidly transports 
larger (0.1--1~mm) particles towards the central star.

These two processes have a dramatic effect on debris production. In a 
standard collisional cascade, mass flows at a roughly constant rate 
from the largest particles to the smallest particles 
\citep[e.g.,][]{wyatt2008,koba2010a,wyatt2011}.  When the gas drags 
0.1--1~mm particles inward, the equilibrium population of these 
particles drops. Collisions among these objects are then less frequent,
depressing the flow of mass to smaller particles. In turn, a smaller mass 
flow rate lowers the cross-sectional area of the swarm and weakens the
infrared excess.

When gas drag and radiation pressure effectively remove small particles 
inside 2~AU, the IR excess drops dramatically. For disks with 0.001--0.01\% 
of the MMSN, the dust luminosity drops by 2--4 orders of magnitude.  
If this remnant disk extends close to the stellar photosphere, particles 
with sizes $\gtrsim$ 1~mm either evaporate or fall onto the photosphere. 
Thus, residual gas disks as dilute as 0.001\% of the MMSN can remove the 
small solids ($<100\,\micron$) that make up 90\% of the cross-sectional 
area of the debris, reducing the IR excess at 5--20~Myr to a level 
consistent with the observational constraints.  Less dilute gas disks 
(0.01\% of the MMSN) can reduce the population of small grains $<100\,\micron$ 
indirectly, by removing the 1--10~mm solids that generate them through 
collisions.  More dilute gas disks are less efficient in removing solids 
$> 10\,\micron$ and therefore less able to limit the IR excess.

We conclude that a residual gas disk in the range 0.001--1\% of the MMSN 
that persists 
for $\sim 10$\,Myr is dilute enough to allow giant impacts to assemble 
terrestrial planets in classical planet formation, but dense enough to 
minimize the observable IR excess produced by the resulting debris.  

%%If disk dissipation {\bf instead} rapidly reduces the gas column density 
%%from 1\% of the MMSN to much less than 0.001\% of the MMSN, gas drag 
%%{\bf would play no role in the giant impact phase, and debris from giant impacts
%%would generate} an observable IR excess.

\subsection{Observational Constraints on Residual Gas Disks}
\label{sec: disc-resid}

%{\bf SK made some mods below}

Current observational constraints on the surface density of residual 
circumstellar gas disks cannot exclude our
picture for reducing 
IR excesses around 10~Myr old solar-type stars. 
Constraints on the residual gas content of disks from {\it in situ} 
diagnostics are model-dependent \citep{gorti2004} and not 
highly restrictive in this context.  For example, analyses 
of {\it Spitzer} IRS emission line diagnostics for a sample 
of FEPS targets with little evidence for warm debris disks 
place a formal limit on the gas surface density 
$\Sigma_g < 0.01$\% of MMSN beyond 1~AU \citep{pascucci2006}.
If this upper limit were to extend inside 1\,AU, the surface density 
would be low enough to allow giant impacts 
in the classical planet formation picture
\citep[e.g.,][]{kominami2002,kominami2004} and 
large enough to allow the removal of small particles by gas drag.

%Using {\it in situ} diagnostics to constrain the amount of residual
%disk gas requires observationally verified thermal-chemical models
%of residual gaseous disks to reliably convert diagnostics (measurements
%or upper limits) into gas column densities.  While models exist
%(e.g., Gorti \& Hollenbach 2004), their predictions await observational
%verification.  Nevertheless, the observational upper limits for
%{\it Spitzer} IRS emission line diagnostics, when interpreted with
%such models, correspond to gas column density upper limits of $\sim
%0.01$\% of MMSN at 1 AU (Pascucci et al.\ 2006). Smaller gas column
%densities ($\sim 0.001$\% MMSN; Fig. 7) can remove small solids
%from the terrestrial planet region of the disk and erase the debris
%disk signature.

%Measuring stellar accretion is an alternate way to search for evidence 
%of a residual gaseous disk. Among systems with opaque disks, average
%accretion rates decrease from $\sim 10^{-8} \Msunperyr$ at 1--3 Myr to
%a few $\times 10^{-10}$ \msunyr\ at 10~Myr, with a factor of 10 spread
%at all ages \citep[e.g.,][]{sicilia2005,sicilia2006,ingleby2013,
%ingleby2014,anton2014}. For standard disk models, a factor of $\sim$ 
%100 decline in the mass accretion rate corresponds to a factor of 
%$\sim$ 100 decline in $\Sigma_g$ \citep[e.g.][]{hart1998}. If the
%high accretion rate sources have MMSN disks, 

Measuring stellar accretion is an alternate way to search for evidence 
of a residual gas disk. While accretion indicates that residual gas 
is present, it does not directly measure $\Sigma_g$.
To obtain a rough scaling between stellar accretion rate $\Mdotstar$
and $\Sigma_g$, we can consider the average accretion 
rates of classical T Tauri stars at 1--10 Myr age. 
Stellar accretion rates decrease from $\sim 10^{-8} \Msunperyr$ 
at the age of Taurus-Auriga (few Myr) to $\sim 4\times 10^{-10}\Msunperyr$
at 10 Myr \citep{sicilia2006}, i.e., only a factor of 25. 
If the fractional decrement in $\Mdotstar$ indicates a 
similar reduction in $\Sigma_g$, an accretion rate of 
$\sim 4\times 10^{-10}\Msunperyr$ corresponds to a surface density 
at 1\,AU of $4\gpersqcm$ if the initial state is an $\alpha$-disk 
with a surface density of $100\gpersqcm$ at 1 AU \citep{hart1998} 
or $80\gpersqcm$  if the initial state is the MMSN with a surface 
density of $2000\gpersqcm$ at 1 AU \citep{weiden1977b}. 
Non-accreting sources, with presumably lower $\Sigma_g$, 
make up more than 90\% of young stars at 5--10 Myr age 
\citep{fedele2010}. 
%
%Thus, even the ``typical'' accretion rate at 10 Myr 
%corresponds to a large reservoir of disk gas,
%several orders of magnitude larger than the limiting gas column
%densities that can remove small solids from the terrestrial planet
%region of the disk ($\gtrsim 0.001$\% MMSN; Fig. 7).
%
%Sources classified as ``non-accreting'' (weak T Tauri stars) have, by definition,
%lower accretion rates than classical T Tauri stars, and it is commonly assumed that
%they also have small to negligible gaseous reservoirs. However true
%upper limits on their accretion rates have only recently been
%investigated in some detail. 

Based on the $\Mdotstar$--$\Sigma$ scaling relation, 
it appears that we need to probe very low accretion rates 
$\sim 10^{-13}-10^{-12}\Msunperyr$ to reach $\Sigma_g \lesssim$
$ 10^{-5} - 10^{-4}$ of the MMSN and to exclude the possibility 
that 5--10 Myr old disks contain enough residual gas to erase a 
debris disk signature.  
For approximately solar mass stars with ages of 5--10 Myr, 
chromospheric emission limits our ability to measure stellar accretion rates 
below $\sim 10^{-10}\Msunperyr$ using $U$-band excesses \citep{ingleby2011b} or 
hydrogen emission line fluxes \citep{manara2013}. H$\alpha$ emission
line widths \citep{natta2004} and line profile analyses 
\citep{muze1998a,muze1998b} 
can probe lower accretion rates \citep[e.g.,][]{muze2000,manara2013}.

From the shape and velocity extent of the H$\alpha$ line profile \citep{white2003,
natta2004}, \citet{riv2015} derive a broad range of accretion rates for 11 sources 
in the 8 Myr old $\eta$ Cha group. Six sources have 
$\Mdotstar \gtrsim 10^{-10}\Msunperyr$ from at least one spectrum; the other five 
sources have $\Mdotstar \approx 10^{-12} - 3\times 10^{-11}\Msunperyr$. While it 
may not be appropriate to apply the \citet{natta2004} relation derived for low mass 
stars to a set of solar-type stars, these accretion rates are close to the rates 
required for gas drag to eliminate the IR excesses of debris disks.

There are multiple prospects for detecting a dilute reservoir of gas in 
solar-type stars. CO fundamental emission \citep[e.g.,][Doppmann, Najita, 
\& Carr 2016]{najita2003,salyk2011}
and UV transitions of molecular hydrogen \citep[e.g.,][]{ingleby2011a}
probe molecular gas within an AU of young stars.
Among A-type stars with debris disks, several have neutral or ionized 
gas close to the star \citep[e.g.,][and references therein]{hobbs1985,
ferlet1987,welsh1998,redfield2007,montgomery2012,kiefer2014,cataldi2014}. 
Evaporation of comets \citep{lagrange1987,beust1990} and vaporization 
of colliding particles \citep[e.g.,][]{czech2007} might supply some of
this material \citep[see also][]{kral2016}.

\subsection{Testing the Possibilities}
\label{sec: disc-test}

In our analysis, we have considered four main options to resolve the 
discrepancy between the frequencies of warm debris disks $f_d$ and 
Earth-mass planets $f_p$ among solar-type stars. Despite clear gaps in 
our understanding of the physics of terrestrial planet formation, there 
is no obvious way to modify the theories to limit the detectability of 
warm debris at 5--20~Myr.  Uncertainties in our ability to relate IR 
excesses to debris from rocky planet formation are unlikely to change 
$f_d$.  Although overestimating the observed frequency of Earth-mass 
planets is a plausible cause for the difference between $f_d$ and $f_p$, 
the `best' explanation is rapid drift of small particles in a residual 
gaseous circumstellar disk.

Future observations and theoretical investigations can test all four 
explanations. We outline several possibilities.

\begin{itemize}

\item Search for fainter warm excess signatures in large samples of young 
solar-type stars. Current sensitivity limits are not much lower than
theoretical predictions. If warm excesses are simply a factor of 2--3
fainter than predicted, detecting these excesses with more sensitive
surveys \citep[e.g., using MIRI on JWST;][]{wright2003,wells2006,bouchet2015}
may be able to distinguish neat scenarios from quick scenarios. 
For example, placing stricter limits on the level of 16~\mum\ excess
($\lesssim$ 10\% of the stellar photosphere)
would probe initial solid masses in debris of a few percent of MMSN 
(Fig.~\ref{fig: flux-ratio2}).

\item In \kepler\ systems with apparent Earth-mass planets, better limits
on (i) the false-positive rate, (ii) the frequency of multiple planets,
and (iii) the distributions of $e$ and $\imath$ in systems with multiple
Earth-mass planets
would reduce the uncertainties in the fraction of stars that host 
Earth-mass planets inside 1--2~AU.
Theoretical studies into the stability of systems with multiple 
Earth-mass planets would place constraints on the ability of 
these systems to `hide' from radial velocity and transit observations.

\item Hunt for planets around all young (5--20~Myr) solar-type stars.
Several recent studies identify massive planets orbiting several 
pre-main sequence stars \citep[e.g.,][]{mann2016,gaidos2016,david2016,
donati2016,johns2016a,johns2016b}.
%both with and without surrounding debris. 
%with no evidence for accretion from a primordial disk. 
Identifying Earth-mass planets orbiting young stars would constrain the 
timescale of terrestrial planet formation relative to the production of debris. 
%enable new tests of theories for the formation of planets and debris disks.  
The K2 mission \citep{howell2014} might identify more short-period planets 
within several nearby young stellar associations\footnote{Among the 
candidates reported in recent analyses of the K2 data 
\citep[e.g.,][]{foreman2015,vanderburg2016}, 205117205.01 matches the 
position of an M2 pre-main sequence star in the Upper Sco association, 
2MASS J16101473$-$1919095 \citep{luhman2012b}. IR data suggest 
the star has a debris or evolved transitional disk. However, the transit
depth is very uncertain. Clarifying the existence and depth of transits
in this system would begin to place constraints on the frequency of planets
in the youngest stars.}.  TESS \citep{ricker2015,sullivan2015} can search
for planets with a much larger range of orbital periods.

\item Search for evidence of tenuous residual gas disks, 
at the $10^{-5}-10^{-2}$ of MMSN level, around young solar-type 
stars.  Direct detection of gas or robust measurement of very low mass 
accretion rates tests the idea that radiation pressure and aerodynamic 
drag remove the debris of terrestrial planet formation.  In addition to 
surveys with the NIRSPEC or MIRI spectroscopic instruments on JWST 
\citep[e.g.,][]{wright2003,wells2006}, sensitive optical 
\citep[e.g.,][]{xu2016} or radio (e.g., ALMA, VLA) observations might 
reveal low mass gaseous disks in the terrestrial zones of 5--20~Myr old 
solar-type stars.

%{\bf JN: The next one is not as strong as the rest? Bands are difficult 
%to interpret without other info.} 

%\item Among the 20\% to 30\% of solar-type stars with cold debris, search
%for indications of radial outflow of small grains from the terrestrial 
%zone.  Isolating distinct bands of 1--10~\mum\ grains at 4--40~AU 
%\citep[Appendix, Fig.~\ref{fig: drag2}; see also][]{take2001} with 
%JWST or ALMA observations might 
%yield additional insight into grain removal mechanisms.

\item Examine quick and neat modes of planet formation. Although our
current understanding appears to preclude these ideas, it is important
to quantify the ability of planet formation scenarios to assemble planets 
quickly and neatly in the absence of a long-lived gaseous disk. 
Future theoretical calculations of terrestrial planet formation should 
include clear predictions of dust production for comparison with 
observations \citep[e.g.,][]{raymond2011,genda2015a,kb2016a}.

\item Consider the late stages of protostellar disk evolution in more
detail. Current mechanisms for disk dispersal (e.g., photoevaporation
and viscous evolution) do not make firm predictions for the structure
of gaseous material on time scales of 10--100~Myr 
\citep[e.g.,][and references therein]{gorti2015}. For example, including 
the gas from the evaporation of comets or vaporization of colliding 
planetesimals in photoevaporation models would help to constrain the 
long-term evolution of debris and gas in the terrestrial zones of
young stars.

\end{itemize}

\section{SUMMARY}
\label{sec: summary}

In the past decade, detailed analyses of \kepler, \spitz, and 
\wise\ data have established estimates for the frequencies of 
Earth-mass planets and warm dust in the terrestrial zones of
solar-type stars. Although rocky planets are fairly common, the 
expected dusty `signature' of terrestrial planet formation among 
5--20~Myr old stars is a rare phenomenon.  Potential explanations
for the discrepancy include the possibility that
(i) terrestrial planets are much less common than believed,
%
% Deleted this because we examined this in the paper but it wasn't a big effect.
%(ii) small dust grains are absent or radiate very inefficiently, 
%
(ii) planet formation is quicker and/or neater than predicted, and
(iii) some physical mechanism removes warm dust rapidly.  Although 
we cannot preclude some combination of the first two options, 
gas drag and radiation pressure can efficiently eliminate warm 
dust particles from the terrestrial zone when a residual gaseous 
circumstellar disk has a surface density of $> 10^{-5}$ of the MMSN.  
%%$10^{-4} - 10^{-5}$ of the MMSN.  
Current constraints on the gas surface density within 1~AU of 5--10~Myr 
old solar-type stars are consistent with this limit.  Thus, tenuous 
reservoirs of gas may impact our ability to observe the debris 
produced by rocky planet formation. 
%If they do, the frequency of warm debris 
%is not a reliable measure of the frequency of rocky planet formation.
If they do, warm debris is not a reliable signpost of rocky planet formation.

Modifying planet formation theory to resolve the discrepancy is tenable 
if planet formation is much more efficient than currently predicted by 
theory and leaves behind little debris.
This scenario echoes the results of several previous studies 
\citep[e.g.,][]{greaves2010b,najita2014} that attempt to reconcile the 
inventory of solids bound up in known populations of exoplanets with 
the solid masses of protoplanetary disks.  Producing the known exoplanet 
systems from the limited solid reservoirs in protoplanetary (Class I) 
disks requires a planet formation efficiency of roughly 30\%.
%
%efficient planet formation 
%which begins in the massive disks of optically invisible protostars. 
%
%Currently known
%populations of exoplanets require efficiencies of roughly 30\%. 
If there is a substantial population of undiscovered planets or 
if planet formation is a messy process that discards solids by 
producing significant debris,
%accessible only by direct imaging, microlensing, or some other 
%technique, 
%
the required efficiency of planet formation rises.

%{\bf The rarity of warm excesses suggests that terrestrial planet 
%formation is frugal rather than wasteful.}  

In the scenarios we investigate, typically $\sim 80$\% (20\%) 
of the initial solid mass ends up in rocky planets (debris).
To limit debris production, it is advantageous to start the 
planet formation process with an initial mass in solids reasonably 
close to the final mass in stable planets.
Without some process that removes small grains from the disk,
theoretical scenarios 
%with 
that invoke much larger initial mass reservoirs 
%and over-produce planets to explain 
%features observed in the population of \kepler\ planets or in
%the solar system
\citep[e.g.,][]{hansen2012,hansen2013,hansen2015,volk2015,levison2015}
%Despite their success in explaining certain aspects of the 
%terrestrial planet population, these scenarios 
should produce very large IR excesses which violate existing 
constraints from observations of 5--20~Myr solar-type stars. 

%{\bf JN: I thought the last paragraph, as written, lacked a 
%punchy ending to the paper.  Made an attempt to improve below.
%I'm trying to say that theorists should consider what debris 
%their planet formation scenario should produce, because 
%this is an important constraint. As things stand, 
%neater is better. Residual gas could save the day if theory is not 
%neat, but work is needed to show that residual gas actually 
%persists.} 

In this sense, collisional debris can place strong constraints 
on planet formation scenarios. 
%
%Predictions for the rate of dust production throughout 
%the epoch of rocky planet assembly {\bf can, in principle,} 
%enable strong tests of the 
%efficiency and timing of terrestrial planet formation.  
As outlined in \S\ref{sec: debris}--\ref{sec: disc}, 
stringent tests require 
(i) tighter observational constraints on warm dust and residual 
disk gas during the expected epoch of terrestrial planet formation
and (ii) planet assembly simulations which include the effect of 
gas drag in a residual gas disk. 

%As outlined in \S\ref{sec: debris}--\ref{sec: disc}, numerical 
%calculations that predict the rate of dust production throughout 
%the epoch of rocky planet assembly enable strong tests of the 
%efficiency and timing of terrestrial planet formation.  
%\citep[see also][and references therein]{kb2004b,kb2005,raymond2011,
%%genda2015a,kb2016a}. 
%%Models of migration through the terrestrial 
%zone should also attempt to derive the rate of dust production
%from any pre-existing ensemble of rocky solids.  Better observational 
%constraints on warm dust and residual disk gas and simulations 
%including gas drag in a residual gas disk would provide more 
%stringent tests of these scenarios.

%Various observational studies would help us improve our
%understanding of the origin of rocky planets. Among 
%solar-type stars with ages of 5--20~Myr, it is possible 
%to constrain the efficiency and timing of rocky planet 
%assembly with initial estimates for the frequency of 
%rocky planets and better limits on the IR excesses from 
%warm dust {\bf and the amount of residual disk gas}.  
%Although isolating the leftovers 
%of agglomeration processes becomes more challenging as 
%stars age, detecting dust emission from occasional giant 
%impacts and improving estimates for the incidence of 
%rocky planets among older stars will help us to link 
%the physics of protostellar disks to the outcomes of the
%planet formation process.

\acknowledgements

We acknowledge a generous allotment of computer time on the NASA 
`discover' cluster.  We thank G. Herczeg for valuable discussions 
of stellar accretion rates.  Comments from and discussions with 
S. Andrews, J. Carpenter, M. Geller, A. Glassgold, G. Kennedy, 
N. Murray, I. Pascucci, D. Wilner, and an anonymous referee 
improved our presentation.  Portions of this project were 
supported by the {\it NASA Outer Planets Program} through grant 
NNX11AM37G.  The work of JN was performed in part at the Aspen 
Center for Physics, which is supported by National Science 
Foundation grant PHY-1066293.  JN also acknowledges the 
stimulating research environment supported by NASA Agreement 
No. NXX15AD94G to the {\it Earths in Other Solar Systems} program. 

\appendix

\section{Long-term Evolution and Gas Drag in Debris Disks}
\label{sec: app}

\subsection{Analytic Theory for an Extended Disk}
\label{sec: app-ann}

\citet{wyatt2002} and \citet{dom2003} first developed an analytic 
theory to track the long-term evolution of collisional cascades
\citep[see also][]{wyatt2007a,wyatt2007b,heng2010,koba2010a,wyatt2011,
kb2016a}. 
In this model, particles with sizes ranging from \rmin\ to \rmax, 
orbital period $P$, and orbital eccentricity $e$ occupy a single 
annulus with width $\delta a$ at a distance $a$ from a central star 
with mass \mstar, radius \rstar, and luminosity \lstar. Particles 
collide at a rate $n \sigma v$, where $n$ is the number density, 
$\sigma$ is the physical cross-section, and $v$ is the collision 
velocity. Setting the center-of-mass collision energy\footnote{In
our approach, the center-of-mass collision energy is 
$Q_c = \mu v^2 / 2 (m_1 + m_2)$, where the reduced mass 
$\mu = m_1 m_2 / (m_1 + m_2)$} $Q_c$ larger than the binding 
energy \qdstar\ of the largest particles ensures destructive 
collisions which produce debris.  For an initial surface density 
of solids $\Sigma_0$ in the annulus, the collision time is 
\begin{equation}
t_c = {\alpha \rmax\ \rho P \over 12 \pi \Sigma_0 } ~ ,
\end{equation}
where the correction factor is 
$\alpha \approx \alpha_0 (v^2 / \qdstar)^{-n}$
\citep{wyatt2007a,heng2010,koba2010a,wyatt2011}.  
The surface density of material in the annulus then evolves as
\begin{equation}
\Sigma_s(t) = { \Sigma_0 \over 1 + t/t_c } ~ .
\label{eq: sigma-t}
\end{equation}
Other collective properties of the particles -- including the total
mass, cross-sectional area, and relative luminosity -- follow the 
decline of $\Sigma_s$ with time \citep[][KB2016]{wyatt2002,dom2003,
wyatt2007a,koba2010a,wyatt2011}.

Here we expand the analytic theory to a disk\footnote{\citet{kw2010}
developed a semi-analytic theory for extended debris disks around 
A-type stars. For the evolution of IR excesses at specific wavelengths,
our approach is similar; our analytic derivation for the evolution of
the luminosity is new.} extending from an inner radius \ain\ to an 
outer radius \aout. Particles with sizes ranging from \rmin\ to 
\rmax\ have an initial surface density
\begin{equation}
\Sigma_s = \xm \Sigma_0 (a / a_0)^{-p} ~ ,
\end{equation}
where $\Sigma_0$ = 10~\gcms\ is the surface density of solids at 
$a_0$ = 1~AU in the MMSN and \xm\ is a scale factor
\citep{weiden1977b,hayashi1981,kb2008}.  We assume that the 
correction factor $\alpha = \alpha_0 (a / a_0)^{-n}$. Setting 
$P = P_0 (a/a_0)^{3/2}$, the collision time is then 
$t_c = t_0 (a / a_0)^{3/2 + p - n}$, where
\begin{equation}
t_0 = {\alpha_0 \rmax\ \rho P_0 \over 12 \pi \xm \Sigma_0 } ~ .
\label{eq: tc0}
\end{equation}

To derive the evolution of the disk, we assume the particles have
a differential size distribution $N(r) \propto r^{-3.5}$ in each
annulus \citep[e.g.,][]{dohn1969,will1994,tanaka1996b, koba2010a}. 
Other choices lead to similar results \citep{wyatt2011}.  We can 
then relate the cross-sectional area of the swarm to the mass 
in each annulus:
\begin{equation}
dA = { 3 \over 4 \rho } \left ( { 1 \over \rmin\ \rmax } \right )^{1/2} dM ~ .
\end{equation}
Setting $dM = \Sigma_s 2 \pi a da $, $\atil = a / a_0$ and $d\atil = da / a_0$,
%\begin{equation}
%dA = { 3 \pi \Sigma a da \over 2 \rho } \left ( { 1 \over \rmin \rmax } \right )^{1/2} ~ .
%\end{equation}
the general result of the analytic model in eq.~(\ref{eq: sigma-t})
and our relation for $t_c$ yields:
\begin{equation}
dA(t) = { 3 \pi \xm \Sigma_0 a_0^2 \atil^{5/2-n} d\atil \over 2 \rho } 
\left ( { 1 \over \rmin \rmax } \right )^{1/2} 
\left ( \atil^{-n+p+3/2} + t / t_0 \right )^{-1} ~ .
\label{eq: dAt}
\end{equation}

To relate this evolution to an observable quantity, we set the
relative luminosity $dL(t) / \lstar = dA(t) / 4 \pi a^2$ = 
$dA(t) / 4 \pi \atil^2 a_0^2$. We then have:
\begin{equation}
{ dL(t) \over \lstar } = { 3 \xm \Sigma_0 \atil^{1/2-n} d\atil \over 8 \rho } 
\left ( { 1 \over \rmin \rmax } \right )^{1/2} 
\left ( \atil^{-n+p+3/2} + t / t_0 \right )^{-1} ~ .
\label{eq: dLt}
\end{equation}
Integrating eqs.~(\ref{eq: dAt}--\ref{eq: dLt}) over $\atil$ yields 
the time evolution of the cross-sectional area and relative luminosity for 
material between $\atilin = \ain\ / a_0$ and $\atilout = \aout / a_0$.

Applying this theory to a real disk requires specifying parameters for
the particles (\rmin, \rmax, and $\rho$) and the disk ($\alpha_0$, $n$, 
\xm, and $p$). Radiation pressure from the central star ejects particles
with radii smaller than \rmin\ \citep{burns1979}; here, we set \rmin\ = 
1~\mum. For equal mass objects with relative velocity $v \approx e v_K$, 
the collision energy is $Q_c = v^2 / 8$. The binding energy is
\begin{equation}
\qdstar = Q_b r^{\beta_b} + Q_g \rho_p r^{\beta_g}
\label{eq: qd}
\end{equation}
where $Q_b r^{\beta_b}$ is the bulk component of the binding energy and
$Q_g \rho_g r^{\beta_g}$ is the gravity component of the binding energy
\citep[e.g.,][]{benz1999,lein2008,lein2009}. Adopting parameters appropriate
for rocky objects and setting $Q_c \approx \qdstar$ yields (KB2016):
\begin{equation}
r_{c,max} \approx 300
\left ( {e \over 0.1} \right )^{1.48}
\left ( {v_K \over {\rm 30~\kms} } \right )^{1.48}
\left ( {\rho \over {\rm 3~\gcmc} } \right )^{-0.74}
\left ( {Q_g \over {\rm 0.3~erg~g^{-1}} } \right )^{-0.74} ~ {\rm km} ~ .
\label{eq: rc-max}
\end{equation}
Based on several simulations of planet formation in the terrestrial zone,
$e \approx$ 0.1 is a reasonable choice 
\citep[e.g.,][]{weiden1997b,kb2006,raymond2007a,chambers2013}.  
Setting \rmax\ = 300~km, the collision time at 1~AU is:
\begin{equation}
t_0 \approx 2.4 \times 10^5 ~ \alpha_0 ~ \xm^{-1} ~
\left ( {r_{max} \over {\rm 300~km} } \right )
\left ( {\Sigma_0 \over 10~\gcms\ } \right )^{-1}
\left ( {P_0 \over {\rm 1~yr}} \right ) ~ {\rm yr} ~ .
\label{eq: tc0n}
\end{equation}

For a standard protostellar disk with $\Sigma_s \propto a^{-p}$, we adopt 
$p$ = 3/2 \citep[e.g.][]{birn2010, windmark2012, testi2014}.  In the 
standard analytic model, $\alpha \propto (v^2 / \qdstar)^{-n}$ with 
$n$ = 5/6 \citep{wyatt2007a, wyatt2007b,koba2010a}. However, comprehensive numerical simulations suggest 
$n$ = 1 (KB2016). Here, we consider $\alpha_0$ = 1 and $n$ = 0 
($v$ = constant throughout the disk) or $n$ = 1 ($e$ = constant throughout 
the disk).  Adopting $\ttil = t/t_0$, the luminosity evolution is then
\begin{equation}
{L(t) \over \lstar } = {\xm \Sigma_0 \over 4 \rho } 
\left ( { 1 \over \rmin \rmax } \right )^{1/2} 
\left [ 
{\rm tan^{-1}} \left ( { \atilout^{3/2} \over \sqrt{\ttil} } \right ) -
{\rm tan^{-1}} \left ( { \atilin^{3/2} \over \sqrt{\ttil} } \right )
\right ] 
\ttil^{-1/2} ~ ,
\end{equation}
for $n$ = 0 and
\begin{equation}
{L(t) \over \lstar } = { 3\sqrt{2} \xm \Sigma_0 \over 32 \rho }
\left ( { 1 \over r_{min} r_{max} } \right )^{1/2} 
\ttil^{-3/4}
\Lambda(\atil,\ttil) 
\end{equation}
for $n$ = 1, where
\begin{equation}
\Lambda(\atil,\ttil) = 
\left[\ln\left(\frac{\atil+\sqrttt+\sqrt{2\atil\sqrttt}}%
{\atil+\sqrttt-\sqrt{2\atil\sqrttt}}\right) 
+2\tan^{-1}\left(\sqrt{2\atil/\sqrttt}+1\right)
+2\tan^{-1}\left(\sqrt{2\atil/\sqrttt}-1\right)
\right]_{\atilin}^{\atilout} ~ .
\end{equation}

At $t$ = 0, solutions for any $n$ yield the same initial luminosity:
\begin{equation}
{L_{init} \over \lstar } = {\xm \Sigma_0 \over 4 \rho } 
\left ( { 1 \over \rmin \rmax } \right )^{1/2} 
\left ( \atilin^{-3/2} - \atilout^{-3/2} \right ) ~ .
\label{eq: linit1}
\end{equation}
In terms of our adopted parameters:
\begin{equation}
{L_{init} \over \lstar } \approx 0.5 \xm 
\left ( { \rmin \over 1~\mum\ } \right )^{-1/2} 
\left ( { \rmax \over {\rm 300~km} } \right )^{-1/2} 
\left ( { \ain \over {\rm 0.1~AU} } \right )^{-3/2}
\left [
1 - \left ( { \ain \over \aout } \right)^{3/2}
\right ] ~ .
\label{eq: linit2}
\end{equation}
When $\xm\ \gtrsim 2 \times 10^{-4}$, the initial luminosity
of the debris disk exceeds our nominal detection limit,
$\ldlstar \gtrsim 10^{-4}$.
At late times, we derive a single expression when $n$ = 0 or $n$ = 1:
\begin{equation}
{L_d(t \gg t_0) \over \lstar } \simeq {(2n+1) \xm \Sigma_0 \over 4 \rho } 
\left ( \atilout^{-n+3/2} - \atilin^{-n+3/2} \right )
\left ( { 1 \over \rmin \rmax } \right )^{1/2} 
\ttil^{-1} ~ .
\label{eq: ldlate1}
\end{equation}
As in the standard model for a single annulus, the luminosity declines
linearly with time when $t \gg t_0$.  For our adopted parameters and 
$n$ = 0:
\begin{equation}
{L_d(t \gg t_0) \over \lstar } \approx 0.04 \xm
\left ( { \rmin \over 1~\mum\ } \right )^{-1/2} 
\left ( { \rmax \over {\rm 300~km} } \right )^{-1/2} 
\left ( { \aout \over {\rm 2~AU} } \right )^{-3/2}
\left [ 1 - \left ( { \aout \over \ain } \right )^{3/2}
\right ]
\ttil^{-1} ~ .
\label{eq: ldlate2}
\end{equation}
With \xm\ = 1 and $t$ = 10~Myr, $L_d / \lstar \approx 10^{-3}$. 
The relative luminosity reaches the nominal detection limit of 
$L_d / \lstar \approx 10^{-4}$ at roughly 100~Myr.

\begin{figure}
\includegraphics[width=6.5in]{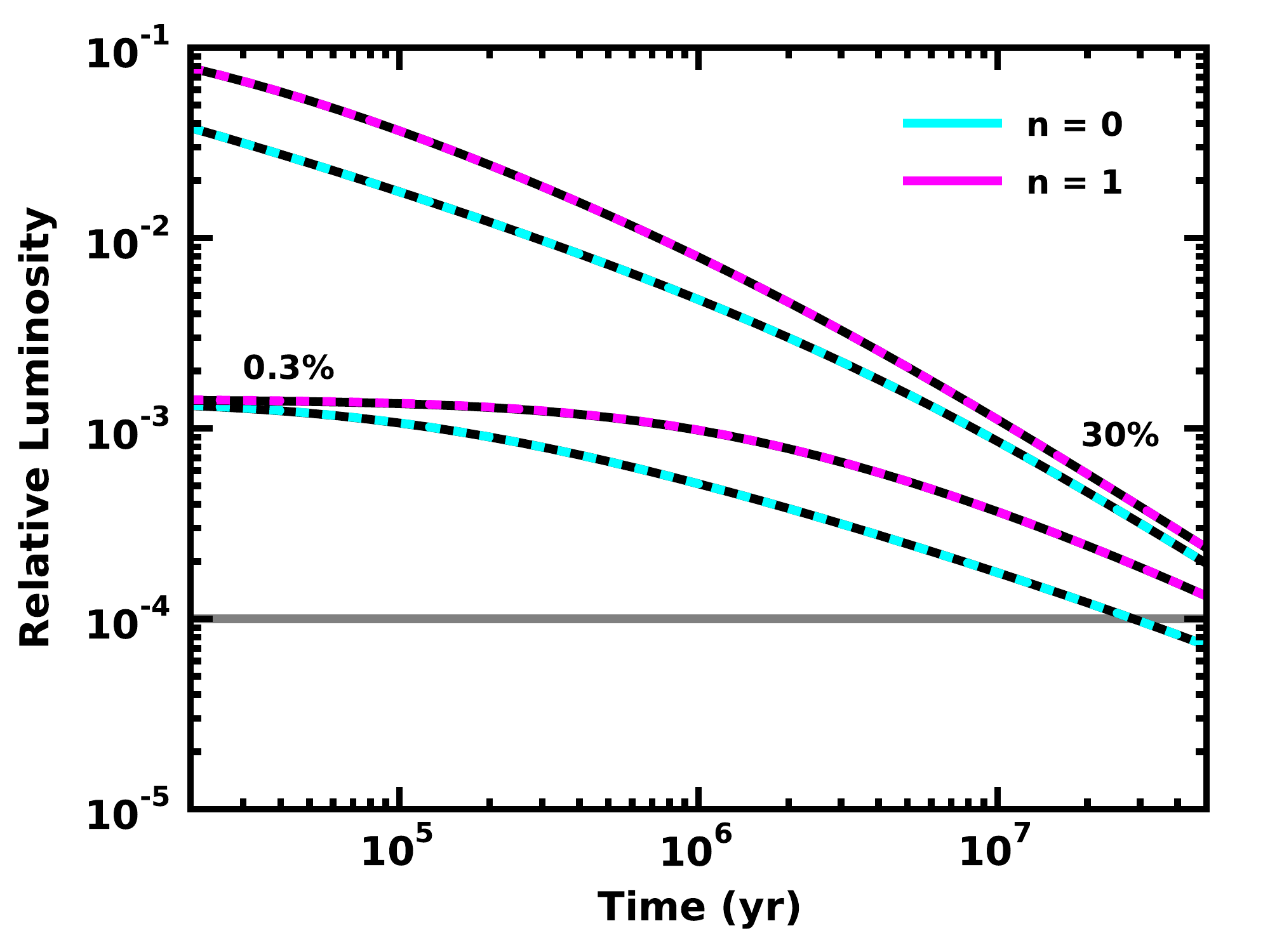}
\vskip 3ex
\caption{
Comparison of numerical (dashed black lines) and analytical
(dashed cyan or magenta lines) results for the time evolution 
of debris with \rmin\ = 1~\mum, \rmax\ = 300~km in a disk with 
\xm\ = 0.3 (upper pair of curves) and \xm\ = 0.003 (lower pair
of curves) extending from 0.1~AU to 2~AU. The legend indicates 
the value of $n$ in the expression 
$\alpha = \alpha_0 (v^2 / \qdstar)^{-n}$ with $\alpha_0$ = 1.
The numerical solution matches the analytic solution to better
than 1\% at $t \lesssim$ 1--10~Myr and to better than 0.1\% at
$t \gtrsim$ 10~Myr.
The grey horizontal line establishes an approximate lower limit
on the luminosity detectable with modern instruments.
\label{fig: an2d}
}
\end{figure}

Fig.~\ref{fig: an2d} illustrates results for a disk with $a_{in}$ 
= 0.1~AU and $a_{out}$ = 2~AU. The two sets of curves compare the 
analytic model with numerical results derived from dividing the 
disk into 101 annuli and summing $dL(t) / \lstar$ (eq.~\ref{eq: dLt}) 
from each annulus. There is superb agreement between the two 
approaches. For any time, the difference between the analytic and
numerical results for $n$ = 0 or $n$ = 1 is less than 0.1\%.
At 10~Myr, all of the model curves lie above the approximate 
detection limit of $L_d / \lstar \approx 10^{-4}$.

To derive the predicted evolution of disk fluxes at specific wavelengths,
we assume all particles radiate as blackbodies at the equilibrium
temperature, $dF(t) = dA(t) B_{\nu}(T)$, where $T = 280~(a/a_0)^{-1/2}$~K. 
Although $dF(t)$ might be integrable, the result is very messy. Given the
good agreement between the analytical and numerical results for 
$L_d(t) / \lstar$, we integrate $dF(t)$ numerically.

\subsection{Gas Drag in a Depleted Disk}
\label{sec: app-gd}

As outlined in the main text, drag within a remnant gaseous disk 
can rapidly remove small particles from a debris disk around 10~Myr 
old stars. To quantify this drift, we adopt a model gaseous disk 
\citep[e.g.,][]{ada76,weiden1977a,kh1987,chiang1997,take2001,
raf2004,chiang2010,youdin2013,armitage2013} with surface density, 
$\Sigma_g = \Sigma_{g,0} (a / {\rm 1~AU} )^{-p}$, 
midplane temperature $T_g = T_{g,0} (a / {\rm 1~AU})^{-q}$, and
vertical scale height $H/a = H_0 (a / {\rm 1~AU})^s$. The midplane
density is $\rho_g = \Sigma_g / 2 H$.  Setting the sound speed
$c_s^2 = (\gamma k_B T_g / \mu m_H)$ -- where 
$\gamma$ is the ratio of specific heats,
$k_B$ is Boltzmann's constant,
$\mu$ is the mean molecular weight, and
$m_H$ is the mass of a hydrogen atom, the gas pressure in 
the midplane is $P_g = \rho c_s^2 / \gamma$. The gas has 
mean-free-path $\lambda = \mu m_H c_s / \Omega \Sigma_g \sigma_{H_2}$ 
and viscosity $\nu = \lambda v_t \rho_g / 3$, where 
$\sigma_{H_2} \simeq 10^{-15}$~cm$^2$ is the collision 
cross-section for H$_2$, and 
$v_t^2 = (8 k_B T_g / \pi \mu m_H)$ is the thermal velocity.

For the calculations in this paper we adopt $p$ = 1, $q$ = 0.5, 
$H_0$ = 0.03, $s$ = 1/8, $\gamma$ = 1.4, and $\mu$ = 2.0. Other
choices for the exponents -- $p$ = 0.5--1.5, $q$ = 0.3--0.7,
and $s$ = 0.0--0.3 -- have modest impact on the results.
In all configurations, the ratio of the mass density of solids
to the mass density of gas is small.

Particles with radii $r \le 9 \lambda / 4$ are in the 
Epstein regime, where the drag force is 
$F_d = \pi r^2 \rho v_d (v_t^2 + v_d^2)^{1/2}$. As in
\citet{take2001}, the term $(v_t^2 + v_d^2)^{1/2}$ provides an 
estimate of the drag force in the subsonic and supersonic regimes.
For larger particles, all drift is in the subsonic regime, 
where the drag force is 
$F_d = C_d \pi r^2 \rho v_d^2 / 2$ and $C_d$ is a drag 
coefficient which depends on the Reynolds number
$Re = 2 \rho v_d r / \nu$
\citep[e.g., eqs. 8a--8c in][]{weiden1977a}.

The main parameters in this model are the gravity of the central 
star $g = G \mstar / a^2$, the residual gravity 
$\Delta g = \rho_g^{-1} dP/da $, and the stopping time
$t_s = m v_d / F_d$ \citep{ada1976,weiden1977a,take2001}. 
The gas orbits at a slightly smaller velocity than the 
local circular velocity. Defining $\eta = -\Delta g / g$, 
the gas velocity is
\begin{equation}
v_g = v_K (1 - \eta)^{1/2} ~ .
\end{equation}
The maximum velocity of drifting particles is roughly:
\begin{equation}
v_{d,max} \approx \eta v_K / 2 ~ .
\end{equation}

In an optically thin disk, radiation pressure drives small
particles radially outward \citep[e.g.,][]{burns1979,take2001}.
Defining $\beta$ as the ratio of the radiation pressure force
to the gravitational force:
\begin{equation}
\beta = { 3 \lstar Q_{pr} \over 16 \pi c G \mstar r \rho_s} ~ ,
\end{equation}
where $c$ is the speed of light. Although \citet{take2001} also
consider the impact of Poynting-Robertson drag, this drag has
little impact when the collision time is short. Thus, we ignore
Poynting-Robertson drag in this discussion. 

\citet{weiden1977a} and \citet{take2001} divide the drift velocity 
of solid particles relative to the gas into a radial component 
$\vrad$ and an azimuthal component $v_\theta$, with 
$v_d^2 = \vrad^2 + v_\theta^2$.  Defining the angular velocity
$\Omega = (G \mstar / a^3)^{1/2}$ and setting $\tau_s = t_s \Omega$, 
the two components of the drift are
\begin{equation}
\vrad = (\eta - \beta) v_K / (\tau_s + \tau_s^{-1}) ~ 
\label{eq: vr}
\end{equation}
and 
\begin{equation}
v_\theta = \tau_s \vrad\ / 2.
\label{eq: vtheta}
\end{equation}

Three parameters -- $\beta$, $\eta$, and $\tau_s$ -- divide drift into
four major regimes.  Large particles poorly-coupled to the gas 
($\tau_s \gg 1$) follow circular orbits with $\vrad\ \approx$ 0 and 
$v_\theta \approx \eta v_K$.  Intermediate size particles weakly 
coupled to the gas ($\tau_s \approx 1$) have maximal drift velocities
and somewhat smaller angular velocities relative to the gas.  Small 
particles ($\tau_s \ll$ 1) are entrained in the gas.  When 
$\eta \gtrsim \beta$ ($\eta \lesssim \beta$) for small objects, 
\vrad\ is positive (negative); particles drift inward (outward).

To identify these regimes in gaseous disks at 1 AU, we develop an 
iterative technique to solve eqs.~(\ref{eq: vr}--\ref{eq: vtheta}).
For each particle size, the algorithm adopts a drag regime, derives
$t_s$ and $\tau_s$, infers \vrad\ and $v_\theta$, and then verifies
the drag regime. This process repeats until the final drag regime 
is identical to the initial drag regime.  When particles are small 
enough to experience Epstein drag, the algorithm loops through an 
initial iteration to derive a consistent $v_d$ for the subsonic and 
supersonic regimes. Comparisons with results in \citet{weiden1977a} 
and \citet{take2001} confirm the accuracy of our approach.

\begin{figure}
\includegraphics[width=6.5in]{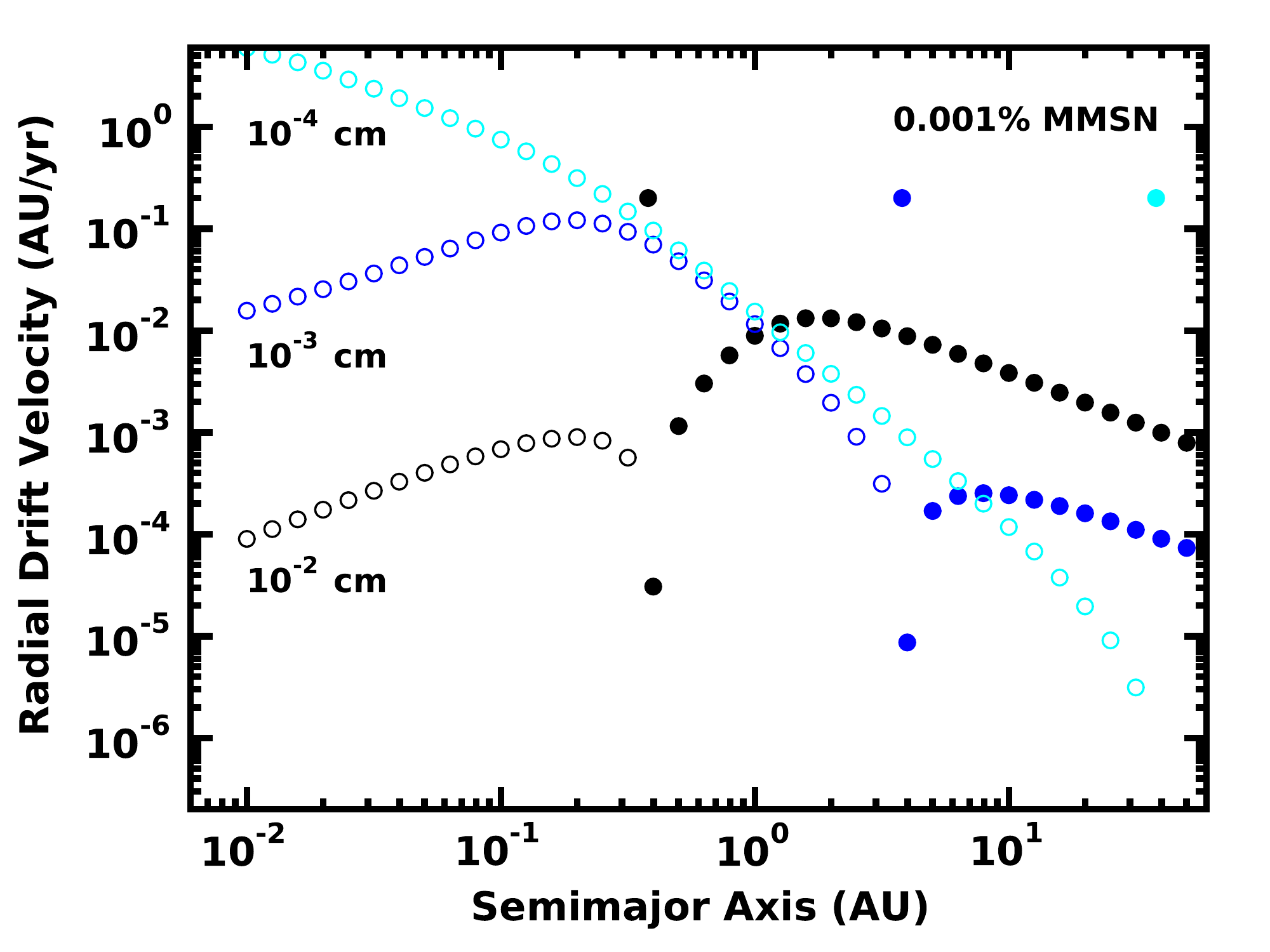}
\vskip 3ex
\caption{
As in Fig.~\ref{fig: drag1} for the drift of 1, 10, and 
100~\mum\ particles as a function of distance from the central 
star. 
\label{fig: drag2}
}
\end{figure}

Fig.~\ref{fig: drag2} illustrates the variation in drift velocity with
semimajor axis for particles with sizes of 1-100~\mum~ in a gaseous
disk with 
a surface density of 0.001\% of the MMSN. Close to the star, radiation
pressure drives particles to larger $a$ (open circles). Far from the
star, the gas drags particles inward (filled circles). At some 
intermediate $a$, small particles find an equilibrium distance $a_{eq}$ 
where radiation pressure balances gas drag. For the disk model adopted 
here, this equilibrium is at
\begin{equation}
a_{eq} \approx { 40~\mum \over r } ~ {\rm AU} ~ .
\label{fig: aeq}
\end{equation}

In solar-type stars with \rmin\ $\approx$ 1~\mum, small particles with 
$r \approx$ 1--10~\mum\ produce most of the IR excess. When the disk's 
surface density is 0.001\% of the MMSN, these small particles have
$a_{eq} \gtrsim$ 4~AU and $T_d \lesssim$ 140~K. These particles are
then too cold to produce a warm debris disk.

%\bibliography{sfpl}
\bibliography{ms.bbl}

\clearpage

\end{document}